\documentclass[10pt,aps,prb,twocolumn,nofootinbib,superscriptaddress]{revtex4-2}

\usepackage[utf8]{inputenc}
\usepackage{graphicx}
\usepackage{color,soul}
\usepackage{bm}
\usepackage{amsmath}
\usepackage{amssymb}
\usepackage{braket}
\usepackage[dvipsnames,svgnames,table]{xcolor}
\usepackage{fancyhdr}
\usepackage[english]{babel}
\usepackage{times}
\usepackage{soul}
\usepackage{hyperref}

\hypersetup{
    unicode=true,		% non-Latin characters in Acrobat's bookmarks
    pdftoolbar=true,		% show Acrobat's toolbar?
    pdfmenubar=true,		% show Acrobat's menu?
    pdffitwindow=false,		% window fit to page when opened
    pdfstartview={FitH},	% fits the width of the page to the windowí
    pdfauthor={ELM},		% author
    colorlinks=true,		% false: boxed links; true: colored links
    linkcolor=NavyBlue,		% color of internal links (change box color with linkbordercolor)
    citecolor=Maroon,		% color of links to bibliography
    filecolor=NavyBlue,		% color of file links
    urlcolor=NavyBlue		% color of external links
}

\newcommand{\tf}[1]{\mathrm{#1}}

\begin{document}

\title{Current-induced molecular dissociation: Topological insulators as robust reaction platforms}

\author{Erika L. Mehring}
\affiliation{Instituto de F\'{i}sica Enrique Gaviola (IFEG-CONICET) and FaMAF, Universidad Nacional de C\'{o}rdoba, Argentina}

\author{Amparo Figueroa}
\affiliation{Instituto de F\'{i}sica Enrique Gaviola (IFEG-CONICET) and FaMAF, Universidad Nacional de C\'{o}rdoba, Argentina}
\affiliation{Department of Physics, The Grainger College of Engineering, University of Illinois Urbana-Champaign,
Urbana, Illinois 61801, USA}

\author{Matias Berdakin}
\affiliation{Consejo Nacional de Investigaciones Cient\'{i}ficas y T\'{e}cnicas (CONICET), Instituto de Investigaciones en Fisicoqu\'{i}mica de C\'{o}rdoba (INFIQC), X5000HUA, C\'{o}rdoba, Argentina}
\affiliation{Universidad Nacional de C\'{o}rdoba, Facultad de Ciencias Qu\'{i}micas, Departamento de Qu\'{i}mica Te\'{o}rica y Computacional, X5000HUA, C\'{o}rdoba, Argentina}
\affiliation{Universidad Nacional de C\'{o}rdoba, Centro L\'{a}ser de Ciencias Moleculares, X5000HUA, C\'{o}rdoba, Argentina}

\author{Hern\'{a}n L. Calvo}
\affiliation{Instituto de F\'{i}sica Enrique Gaviola (IFEG-CONICET) and FaMAF, Universidad Nacional de C\'{o}rdoba, Argentina}

\begin{abstract}
The growing interest in topological materials with symmetry-protected surface states as catalytic platforms has sparked the emerging field of \textit{topocatalysis}. As robust transport is one of the key features of topological insulators, here we explore current-induced molecular dissociation in a transport setup. Using the non-equilibrium Green’s function formalism, we compare how the occupancies of bonding and antibonding levels, as well as the associated electronic forces in a diatomic molecule, are affected when the molecule is coupled to either a metallic (graphene) or a topological (Kane-Mele) substrate. We find a greater dissociative capability in the topological substrate than in graphene, a difference mainly attributed to the localized nature of the edge states. The inclusion of vacancy disorder within the substrate further enhances this disparity in the dissociative force. Our findings highlight the role of topological protection in molecular dissociation under non-equilibrium conditions, pointing to new opportunities for robust catalysis in topological materials.
\end{abstract}

\maketitle

\section{Introduction}
\label{sec:intro}

The rise of topological insulators (TIs) has opened a new frontier in condensed matter physics, revealing materials whose bulk remains electrically insulating while their surfaces host metallic states protected by time-reversal symmetry \cite{bernevig2006}. These surface states exhibit spin-momentum locking (spin polarization is rigidly tied to its momentum direction), immunity to backscattering from non-magnetic impurities, and robustness against structural disorder, making them fundamentally different from conventional conductors and semiconductors \cite{hasan2010,ren2016}.

The advent of TIs has fueled expectations for a new generation of quantum and spin-based technologies. Their intrinsic spin-polarized surface (or edge, in 2D TIs) states hold great promise for spintronics, enabling low-dissipation spin transport without the need for external magnetic fields \cite{benitez2020,dang2023}. Furthermore, the unique electronic structure of TIs provides new avenues for controlling charge flow, inspiring device concepts such as the topological field-effect transistor \cite{collins2018}. Beyond electronics, the unusual optical response of TIs suggests potential applications in optoelectronics, for instance in the generation of spin-polarized photocurrents \cite{mciver2012,berdakin2021} and topological switches \cite{ezawa2013}.

\begin{figure}[htbp]
\includegraphics[width=\linewidth]{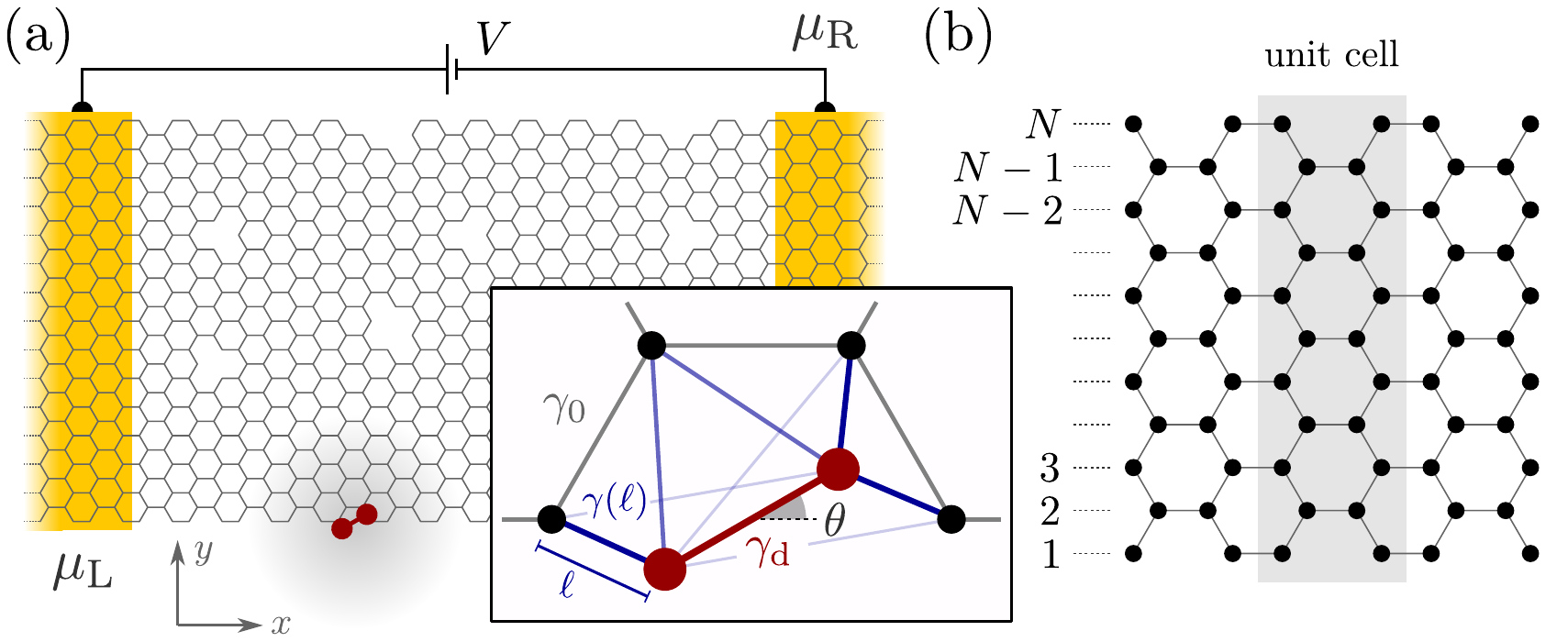}
\caption{(a) Scheme of the catalysis model, based on the interaction between the adsorbate (red circles) and the adsorbent, composed of an armchair graphene nanoribbon with passivated vacancies. The contacts extend on both sides of the sample and are represented by yellow regions. The shaded region around the adsorbate indicates the extent of the coupling with the adsorbent. Inset: Zoom of the molecule-substrate interaction region; with $\gamma_0$ the carbon-carbon hopping, $\gamma_\tf{d}$ the intramolecular coupling, and $\gamma(\ell)$ the hopping term between the atoms of the molecule and the carbon atoms of the substrate. (b) Honeycomb lattice with armchair edge termination. The shaded region denotes the unit cell, containing $2N$ carbon atoms.}
\label{fig:scheme}
\end{figure}

A less explored---but rapidly emerging---area of interest lies in the application of topological materials in catalysis \cite{yang2023,guowei2020,guo-liang2022}. The resilience of topologically protected states to disorder and other perturbations, such as surface passivation, makes them promising candidates for a new generation of catalysts. Recent studies suggest that their topological surface states can influence adsorption and electron-transfer processes, with encouraging results in reactions such as catalytic oxidations \cite{hua2011,xiao2015,hosono2023}, nitroarene reduction \cite{nethravathi2019}, and photocatalytic water splitting \cite{rajamathi2017,li2019,liu2024,xiaoming2023}.

Since transport phenomena represent one of the most promising features of topological insulators, in this work we investigate the role of topological states in the destabilization of molecules interacting with edge currents within a transport setup. In this regard, the effect of electronic currents on the potential dissociation of molecules near topological materials is an incipient subject of study. There are, nevertheless, reports in the literature that are related to exchange effects between a nanoparticle (or molecule) and a topological substrate. For example, several works address the role of spin-momentum locking effect in the properties of a nanomagnet weakly coupled to the helical edge states of a 2D TI in a transport setup \cite{meng2014,arrachea2015,locane2017}. In a related direction, in Ref.~\cite{farias2018} the authors explore the quantum drag force on a metallic nanoparticle moving above a 2D topological substrate.

For our purposes, modeling a full 3D catalytic system in non-equilibrium conditions at the quantum transport level is computationally challenging; therefore, to capture the essential physics while retaining numerical tractability, we restrict our analysis to two-dimensional substrates. In this framework, the catalytic device is modeled as a nanoribbon connected to semi-infinite leads, with the molecule placed near one of its edges to mimic surface adsorption in a real 3D catalyst [see Fig.~\ref{fig:scheme}(a)]. 
In addition, we use a single-particle approach based on tight-binding (TB) models for the composite system. Although neglecting many-body interactions would lead to some quantitative limitation, the obtained results in this work illustrate the role of TIs, which could be extended to a more involved description of the physical system. This minimal setup thus allows us to explore current-induced effects on molecular dissociation while retaining the key differences between standard metallic and topological substrates. 

As a proof of concept, we describe a diatomic molecule as a two-orbital system. Applying a bias voltage across the leads drives a current through the substrate, enabling us to study how charge flow modifies the electronic occupation of molecular levels and the resulting electronic forces. To establish a consistent comparison between metallic and topological catalysts in terms of efficiency and robustness of current-induced molecular destabilization, we consider a graphene ribbon with armchair edge termination. By introducing a spin-orbit coupling (SOC) term, the ribbon becomes a topological insulator described by the Kane-Mele model \cite{kane2005,kane2005a}, opening a topological gap that supports robust helical edge states. In this framework, for both substrates we show that a finite bias can destabilize molecular bonds by promoting electron transfer from bonding to antibonding levels, which counteracts the equilibrium attractive force of the molecular bond, thereby enhancing current-induced dissociation. Comparing trivial graphene with topological substrates, we find that extended states in graphene reduce molecular occupancies of the molecular states as the ribbon width increases, while localized edge states in topological insulators preserve them, sustaining catalytic efficiency. Furthermore, we demonstrate that topological substrates display a remarkable robustness against vacancy disorder, in contrast to the strong sensitivity observed in graphene. These results highlight the potential of topological edge states as efficient and resilient platforms for current-driven catalysis.

\section{Model and Method}
\label{sec:model}

\subsection{Description of the physical system}
 
\noindent \textit{Transport device}.-- To allow the operation of the substrate in a non-equilibrium regime, we consider the inclusion of left (L) and right (R) electrodes, which are modeled as semi-infinite extensions of the ribbon, as it is shown in Fig.~\ref{fig:scheme}(a). These contacts act as electron reservoirs in equilibrium at a temperature $T$ and chemical potentials $\mu_\alpha$, with $\alpha = \{\tf{L},\tf{R}\}$, such that the electron occupancy is given by the Fermi-Dirac distribution function:
\begin{equation}
    f_\alpha(\epsilon)=\frac{1}{1+\exp[(\epsilon-\mu_\alpha)/k_\tf{B} T]}.
\end{equation}
Importantly, the contacts are displayed far away from the molecule, such that the coupling between these subsystems can be neglected. This defines the central system, composed by the interaction region between the molecule and the substrate.

For a non-zero bias voltage $V$ the central region is forced into a stationary non-equilibrium state where a constant electron current flows over time. Here we use symmetric biasing through $\mu_\tf{L} = \mu_0 + eV/2$ and $\mu_\tf{R} = \mu_0 - eV/2$, with $\mu_0$ the equilibrium chemical potential and $e>0$ the magnitude of the electron charge. Additionally, a gate voltage $V_\tf{g}$ can be included to shift the energy levels within the central region without affecting those in the contacts. Although in this work we explicitly take $V_\tf{g}=0$, shifting $\mu_0$ from its natural position at the Dirac point can be interpreted as the inclusion of a local gate potential in the central region. This would be particularly useful, e.g., to control the molecular occupation under equilibrium conditions.\\

\noindent \textit{Hamiltonian model}.-- The full system Hamiltonian, including both substrate (adsorbent) and molecular (adsorbate) contributions, is given by:
\begin{equation}
\hat{H}=\hat{H}_\tf{d}+\hat{H}_\tf{s}+\hat{H}_\tf{int},
\end{equation}
where the substrate Hamiltonian $\hat{H}_\tf{s}$ represents the extended lattice (including the leads), and the molecular Hamiltonian $\hat{H}_\tf{d}$ describes the molecule under study. The coupling term $\hat{H}_{\tf{int}}$ accounts for the interaction between these subsystems.

Throughout this work, we will assume a homonuclear molecule composed of two atoms, each one located at positions $\bm{R}_1$ and $\bm{R}_2$ and described by an $s$-like orbital. The TB Hamiltonian of the diatomic molecule, $\hat{H}_\tf{d}$, can be written as:
\begin{equation}
    \hat{H}_\tf{d} = \sum_{n,\sigma} E_n \hat{d}^\dag_{n,\sigma} \hat{d}_{n,\sigma} - \gamma_\tf{d} \sum_\sigma \left( \hat{d}^\dag_{1,\sigma} \hat{d}_{2,\sigma} + \tf{h.c.} \right),
    \label{eq:ham-dim}
\end{equation}
where $\hat{d}^\dag_{n,\sigma}$ creates an electron with spin $\sigma=\{\uparrow,\downarrow\}$ at site $\bm{R}_n$, with $n=\{1,2\}$. The simplified assumption of non-magnetic components in the molecule enables us to neglect the spin degree of freedom in the onsite energies $E_n$ and hopping term $\gamma_\tf{d}$. The position of the molecule's sites $\bm{R}_n$ can be described as:
\begin{equation}
    \bm{R}_n=\bm{R}_0 +R\left[\cos(\theta+\varphi_n)\bm{e}_x+\sin(\theta+\varphi_n)\bm{e}_y\right],
    \label{eq:position}
\end{equation}
where $R$ is the molecular radius, measured from the center of the molecule at $\bm{R}_0$, and $\theta$ is the angle between the molecule's axis and the transport direction [aligned with the $x$-axis, see Fig.~\ref{fig:scheme}(a)], and we consider the symmetric case where $\varphi_1=0$ and $\varphi_2=\pi$. Additionally, the interaction between the molecule's atoms is modeled through the following decaying exponential:
\begin{equation}
\gamma_\tf{d}=\alpha_\tf{d}\gamma_0\exp \! \left(-b\frac{2R}{a}\right),
\label{eq:hop_dim}
\end{equation}
such that its magnitude and decay factor are controlled by the dimensionless parameters $\alpha_\tf{d}$ and $b$, respectively, referred to the considered graphene substrate, where the carbon-carbon distance is $a = 1.42$ {\AA} and the nearest-neighbor hopping amplitude is $\gamma_0 = 2.7$ eV. This exponential form reflects a covalent bond induced by tunneling, where the decay in the electron probability through a potential barrier is typically characterized by a penetration constant \cite{mehring2024}. In addition, we take $E_1=E_2=0$ so the eigenenergies of the isolated molecule are centered around the substrate's Dirac point, defined at $\epsilon = 0$. Therefore, the intramolecular coupling $\gamma_\tf{d}$ produces symmetric $\ket{+,\sigma}$ and antisymmetric $\ket{-,\sigma}$ eigenstates, where:
\begin{equation}
\ket{\pm,\sigma} = \hat{d}^\dag_{\pm,\sigma} \ket{0} = \frac{\hat{d}^\dag_{1,\sigma} \pm \hat{d}^\dag_{2,\sigma}}{\sqrt{2}} \ket{0}. \label{eq:b-ab}
\end{equation}
with $\ket{0}$ the vacuum state for the isolated molecule. Although these states can be identified with the ``bonding'' and ``antibonding'' molecular states, we will reserve this nomenclature for the projection of the molecule-substrate hybridized states into the subspace of the molecule, as we discuss at the end of the section.

The coupling Hamiltonian between the molecule and the substrate takes the form:
\begin{equation}
    \hat{H}_\tf{int} = - \sum_{n,\sigma} \sum_i \gamma(\bm{r}_i,\bm{R}_n) \left( \hat{c}^\dag_{i,\sigma}\hat{d}_{n,\sigma} + \tf{h.c.} \right),
    \label{eq:ham-int}
\end{equation}
where $\hat{c}^\dag_{i,\sigma}$ creates an electron with spin $\sigma$ at site $i$ of the substrate. Similar to the intramolecular hopping amplitude, the interaction between the molecule's atom $n$, located at $\bm{R}_n$, and the lattice site $i$ of the substrate can be described by:
\begin{equation}
    \gamma(\bm{r}_i,\bm{R}_n)= \alpha_\tf{x} \gamma_0 \exp \! \left(-b \frac{|\bm{r}_i-\bm{R}_n|}{a}\right),
\end{equation}
where the dimensionless parameter $\alpha_\tf{x}$ sets the coupling strength between the molecule and the substrate. Similar TB models were used in the context of quantum dynamical phase transitions to study molecular dissociation in the presence of catalysts \cite{ruderman2015,ruderman2016}.

The substrate or catalyst model consists of a single graphene layer, which can be extended to its topological equivalent, the Kane-Mele model \cite{kane2005}. The latter serves as a paradigmatic testbed to address the non-trivial topological phase $\mathbb{Z}_2$, responsible for the quantum spin Hall effect \cite{konig2007}, characterized by spin-polarized helical edge states. This model successfully describes the underlying physics of 2D topological insulators like silicene or germanene monolayers \cite{liu2011,ezawa2015}. The corresponding Hamiltonian writes:
\begin{equation}
\hat{H}_\tf{s} = -\gamma_0 \sum_{\braket{i,j},\sigma} \hat{c}_{i,\sigma}^\dag \hat{c}_{j,\sigma} + \gamma_1 \sum_{\langle\!\langle i, j \rangle\!\rangle, \sigma} e^{i \phi_{ij,\sigma}} \hat{c}_{i,\sigma}^\dag \hat{c}_{j,\sigma},
\label{eq:ham-sub}
\end{equation}
where $\gamma_1$ characterizes the strength of the intrinsic spin-orbit interaction. The notation $\braket{i,j}$ and $\langle\!\langle i,j\rangle\!\rangle$ indicates that the sums are restricted to first and second neighbors, respectively. The first term in $\hat{H}_\tf{s}$ corresponds to the standard graphene TB Hamiltonian. The second term introduces a complex chiral second-neighbor hopping with spin-dependent phases $\phi_{ij,\sigma}$, whose sign depends on the spin projection $\sigma=\{\uparrow,\downarrow\}$, i.e., $\phi_{ij,\uparrow} = \phi_{ij}$ and $\phi_{ij,\downarrow} = -\phi_{ij}$. This term opens a topological gap in the bulk while giving rise to spin-polarized edge channels. As a result of the spin-orbit coupling, the system hosts helical edge states protected against backscattering. By setting $\gamma_1=0$, the system reduces to standard graphene and enters into a trivial topological phase. In this work, we assume the Semenoff mass term to be zero.

\subsection{Theoretical framework: NEGF approach}
\label{sec:theory}

To compute the studied physical properties, we employ the non-equilibrium Green's function (NEGF) formalism \cite{pastawski1992,jauho1994} which, in the approximation of non-interacting electrons, offers computational efficiency compared to directly solving the Schrödinger equation under non-equilibrium conditions. Within this formalism, the non-equilibrium stationary density matrix can be computed as $\bm{\rho}(t) =-i\hbar\bm{G}^<(t)$, where $\bm{G}^<(t)$ is the lesser Green's function. In the stationary limit this function becomes time-independent, and can be obtained as:
\begin{equation}
\bm{G}^< = \int \frac{\tf{d}\epsilon}{2\pi\hbar} \bm{G}^r(\epsilon) \bm{\Sigma}^<(\epsilon) \bm{G}^a(\epsilon),
\label{eq:lesserG}
\end{equation}
where $\bm{G}^a =\left[\bm{G}^r\right]^\dag$ are the Wigner transforms of the retarded and advanced Green's function matrices, respectively, and $\bm{\Sigma}^<$ is the lesser self-energy. The latter can be written in terms of the Fermi distribution function of the reservoir $\alpha$ as
\begin{equation}
\bm{\Sigma}^<(\epsilon)=2i\sum_\alpha f_\alpha(\epsilon) \bm{\Gamma}_\alpha (\epsilon), 
\end{equation}
where $ \bm{\Gamma}_\alpha = \left(\bm{\Sigma}^a_\alpha-\bm{\Sigma}^r_\alpha\right)/2 i$ represents the escape rate through the $\alpha$ lead, and $\bm{\Sigma}^r_\alpha(\bm{\Sigma}^a_\alpha)$ is the $\alpha$-lead contribution to the total retarded (advanced) self-energy. The general expression for the retarded Green's function is given by:
\begin{equation}
    \bm{G}^r(\epsilon)  = \lim_{\eta \rightarrow 0^+} \left[ \left( \epsilon +i\eta\right)\bm{I}-\bm{H}_\tf{sys}-\bm{\Sigma}^r(\epsilon)\right]^{-1} ,
    \label{eq:green_e}
\end{equation}
where $\bm{H}_\tf{sys}$ denotes the central system Hamiltonian including the substrate, the molecule, and the interaction between them. $\bm{\Sigma}^r = \bm{\Sigma}^r_\tf{L}+\bm{\Sigma}^r_\tf{R}$ is the retarded self-energy correction due to the contacts, whose calculation exploits the translational invariance within the leads, yielding a fast decimation procedure \cite{lopez1985,calvo2013}.

The system's density matrix is thus expressed as:
\begin{equation}
\bm{\rho}=\sum_\alpha \int \frac{\tf{d}\epsilon}{\pi} f_\alpha(\epsilon) \bm{G}^r(\epsilon) \bm{\Gamma}_\alpha(\epsilon) \bm{G}^a(\epsilon) ,
\label{eq:rho-t}
\end{equation}
where, as in Eq.~\eqref{eq:lesserG}, we omit the time argument because we are only interested in the stationary solution. In this expression, an ``occupation kernel'' can be defined as $\bm{\mathcal{K}}^{\rho}_\alpha=\bm{G}^r \bm{\Gamma}_\alpha \bm{G}^a / \pi$, which encapsulates the contribution to the density matrix due to electrons incoming from the $\alpha$ lead. It may also be convenient to separate the equilibrium and non-equilibrium contributions in Eq.~\eqref{eq:rho-t}. By rewriting the distribution functions as $f_\alpha = f_0 + (f_\alpha - f_0)$, we obtain $\bm{\rho} = \bm{\rho}^{\tf{eq}} + \bm{\rho}^{\tf{ne}}$, where the non-equilibrium term is given by:
\begin{equation} 
\bm{\rho}^{\tf{ne}}=\sum_\alpha \int \tf{d}\epsilon [f_\alpha(\epsilon)-f_0(\epsilon)]\bm{\mathcal{K}}^{\rho}_\alpha(\epsilon). 
\end{equation}
When a current passes through atoms, a force acts upon them \cite{todorov2014}. Allowing atomic positions to vary introduces a Hamiltonian that parametrically depends on the atomic coordinates. By using the density matrix, the electronic force can be computed as \cite{bode2011}:
\begin{equation}
F = \tf{Tr}[\bm{\Lambda}\bm{\rho}]= \sum_\alpha \int \tf{d}\epsilon f_\alpha(\epsilon) \mathcal{K}^{F}_\alpha (\epsilon),
\label{eq:force}
\end{equation}
where the matrix $\bm{\Lambda}=-\partial_R\bm{H}$ represents the force operator due to the variation of the molecule's radius, and $\mathcal{K}^{F}_\alpha=\tf{Tr}[\bm{\Lambda}\bm{G}^r\bm{\Gamma}_\alpha\bm{G}^a]/\pi$ is the force kernel associated with the $\alpha$ lead. Once more, this force can be linearly decomposed in terms of equilibrium and non-equilibrium contributions: $F = F^\tf{eq}+F^\tf{ne}$.

Under equilibrium conditions, where no bias is applied and no net force acts on the molecule, the equilibrium force $F^\tf{eq}$ is balanced by other forces not described in our model, which establishes the reference value for analyzing molecular stability. Consequently, in the context of this study, the focus lies on the non-equilibrium contribution $F^\tf{ne}$, as it describes the influence of an applied bias voltage, which is a key driving factor for molecular dissociation in our setup.

Regarding the total Hamiltonian, the electron force $F$ can also be decomposed into a contribution $F_\tf{s}$ arising from the interaction with the substrate through $\hat{H}_\tf{int}$ and a contribution $F_\tf{d}$ associated with the molecule itself through $\hat{H}_\tf{d}$. The former involves the hybridization between the molecular states and the lattice sites, while the latter accounts for the molecular occupancies. For the considered parameters, we find that $F_\tf{s} \ll F_\tf{d}$. This is attributed to the large difference between the non-equilibrium molecular occupancy and that of the substrate sites around the interaction region, as we discuss in the next section [see Fig.~\ref{fig:rho_fza}(a)]. Therefore, throughout the rest of this work, we will restrict our analysis to the intramolecular term $F_\tf{d}$, defined through the following force operator:
\begin{equation} 
\hat{\Lambda}_\tf{d}= -\partial_R \hat{H}_\tf{d} = -\frac{2b}{a}\gamma_\tf{d} \sum_\sigma (\hat{d}^\dagger_{1,\sigma} \hat{d}_{2,\sigma} + \tf{h.c.}). 
\label{eq:fce-d}
\end{equation}
In this convention, a positive value of the force corresponds to a repulsive interaction, favoring molecular dissociation, whereas a negative value indicates an attractive interaction, stabilizing the bond (see Appendix~\ref{app:wb}).

To compute the electronic occupation of the molecular levels and the intramolecular force, we project the full system onto the molecular subspace. In particular, we evaluate the occupancies in a \textit{renormalized} eigenbasis that accounts for the presence of the substrate to avoid the mixing of the molecular levels due to this interaction. To this end, we construct an effective Hamiltonian for the molecular subspace d, spanned by the states $\ket{\pm,\sigma}$ of Eq.~\eqref{eq:b-ab}, that includes the renormalization induced by the substrate subspace s:
\begin{equation}
\tilde{\bm{H}}_\tf{dd}(\epsilon) = \bm{H}_\tf{dd} + \bm{H}_\tf{ds} \left(\epsilon \bm{I} - \bm{H}_\tf{ss}\right)^{-1} \bm{H}_\tf{sd},
\end{equation}
where the diagonal blocks $\bm{H}_\tf{dd}$ and $\bm{H}_\tf{ss}$ are the bare molecular and substrate Hamiltonian matrices of Eqs.~\eqref{eq:ham-dim} and \eqref{eq:ham-sub}, respectively, while the off-diagonal blocks $\bm{H}_\tf{ds}$ and $\bm{H}_\tf{sd}$ are the coupling matrices between these subsystems, representing the interaction Hamiltonian of Eq.~\eqref{eq:ham-int}. By taking $\epsilon$ as a fixed parameter, we diagonalize $\tilde{\bm{H}}_\tf{dd}$ according to:
\begin{equation}
    \tilde{\bm{H}}_\tf{dd}(\epsilon)\bm{u}_m(\epsilon) = \lambda_m(\epsilon) \bm{u}_m(\epsilon),
\end{equation} 
and obtain the substrate-renormalized molecular eigenstates $\bm{u}_m$. Notice that we are not solving here the (non-linear) stationary Schrödinger equation which leads to the poles of the Green's function through the solutions of $\lambda_m(z) = z$, with $z \in \mathbb{C}$. Instead, we use the transformation matrix $\bm{U}$ that diagonalizes $\tilde{\bm{H}}_\tf{dd}$ to compute the molecule's density matrix through the kernel:
\begin{equation}
\tilde{\bm{\mathcal{K}}}_\alpha^\rho(\epsilon) = \bm{U}^\dagger(\epsilon) \bm{\Pi}_\tf{d} \bm{\mathcal{K}}_\alpha^\rho(\epsilon) \bm{\Pi}_\tf{d} \bm{U}(\epsilon),
\end{equation}
where $\bm{\Pi}_\tf{d}=\bm{\Pi}^\dag_\tf{d}$ is the projection operator matrix in the molecular subspace. We will call these molecular states `bonding' (b) and `antibonding' (a), since they essentially preserve the properties of the symmetric and antisymmetric states defined in Eq.~\eqref{eq:b-ab}, for the considered regimes of parameters.

Importantly, as the Kane-Mele substrate leads to a spin-momentum locking effect that clearly impacts the molecular occupation, we need to include the electronic spin degree of freedom in our description of the physical system. In doing so, we define the molecular occupancies through:
\begin{equation}
\rho_m= \frac{1}{2}\sum_\sigma \rho_{m,\sigma},
\label{eq:occ-spin}
\end{equation}
where $\rho_{m,\sigma}$, with $m=\{\tf{b},\tf{a}\}$, represents the occupancy of the $m$ hybridized orbital with spin $\sigma$. For our purposes, this is equivalent to neglecting the spin degree in Eq.~\eqref{eq:ham-sub} (Haldane model, see Ref.~[\onlinecite{haldane1988}]) and calculating the occupancies $\rho_{m,\uparrow}$ and $\rho_{m,\downarrow}$ separately by taking opposite phases $\phi_{ij}$.

Although the description of the hybridized substrate-molecular states is achieved within the single-particle framework, it can be mapped to a more realistic context by translating the bonding and antibonding levels to HOMO and LUMO molecular orbitals, respectively. As we shall discuss, when coupling the molecule to the substrate and adjusting the bias window to include these levels, a transport regime is established where the HOMO level (occupied by $N$ electrons) loses a single electron (oxidation) and, subsequently, another electron occupies the LUMO level (reduction). In this sense, by replacing the reference state $\ket{0}$ by $\ket{N-1}$ in Eq.~\eqref{eq:b-ab}, our formalism describes the occupation/depopulation of the HOMO (bonding) and LUMO (antibonding) molecular orbitals within the fixed charge transition region $N-1 \leftrightarrow N$.

In what follows, we will evaluate both the molecular occupations and related forces to identify scenarios in which the molecule may dissociate due to the substrate's influence under the application of a bias voltage.

\section{Results}

\subsection{Non-equilibrium molecular destabilization}

We initiate our analysis by considering a pristine armchair-edged graphene nanoribbon defined by $N$ carbon atoms along the transversal direction \cite{foatorres2014} [see Fig.~\ref{fig:scheme}(b)]. In particular, we set $N$ under the condition $N=3\ell+2$, with $\ell \in \mathbb{N}$, in order to obtain a metallic ribbon. For this analysis we use $\ell = 25$, resulting in a nanoribbon width of approximately $W=\sqrt{3}(N-1)a/2 \approx 9.3$ nm. The molecular center is located at a distance of $1.5$ {\AA} above the substrate plane, with its bond axis aligned along $\theta=\pi/2$, i.e., parallel to the $y$ direction. In Appendix \ref{app:angle} we extended the analysis to various molecular orientations, though no qualitative differences appear for both the molecular occupancies and the intramolecular force. Additionally, its center is located at the lower edge, i.e., $y_0 = (1-N)\sqrt{3}a/4$, measured from the center of the ribbon. The two atoms are separated by a distance of $1.5$ {\AA} from each other along the bond axis, such that the radius of the molecule is $R = 0.75$ {\AA}. For the topological case ($\gamma_1 \neq 0$), this setup allows us to probe the interaction between the molecule and the localized edge-states. To ensure that the edge states are well defined and spatially separated, the ribbon width must be chosen to be larger than the localization length to avoid hybridization between opposite edges.

\begin{figure}[htbp]
\includegraphics[width=\columnwidth]{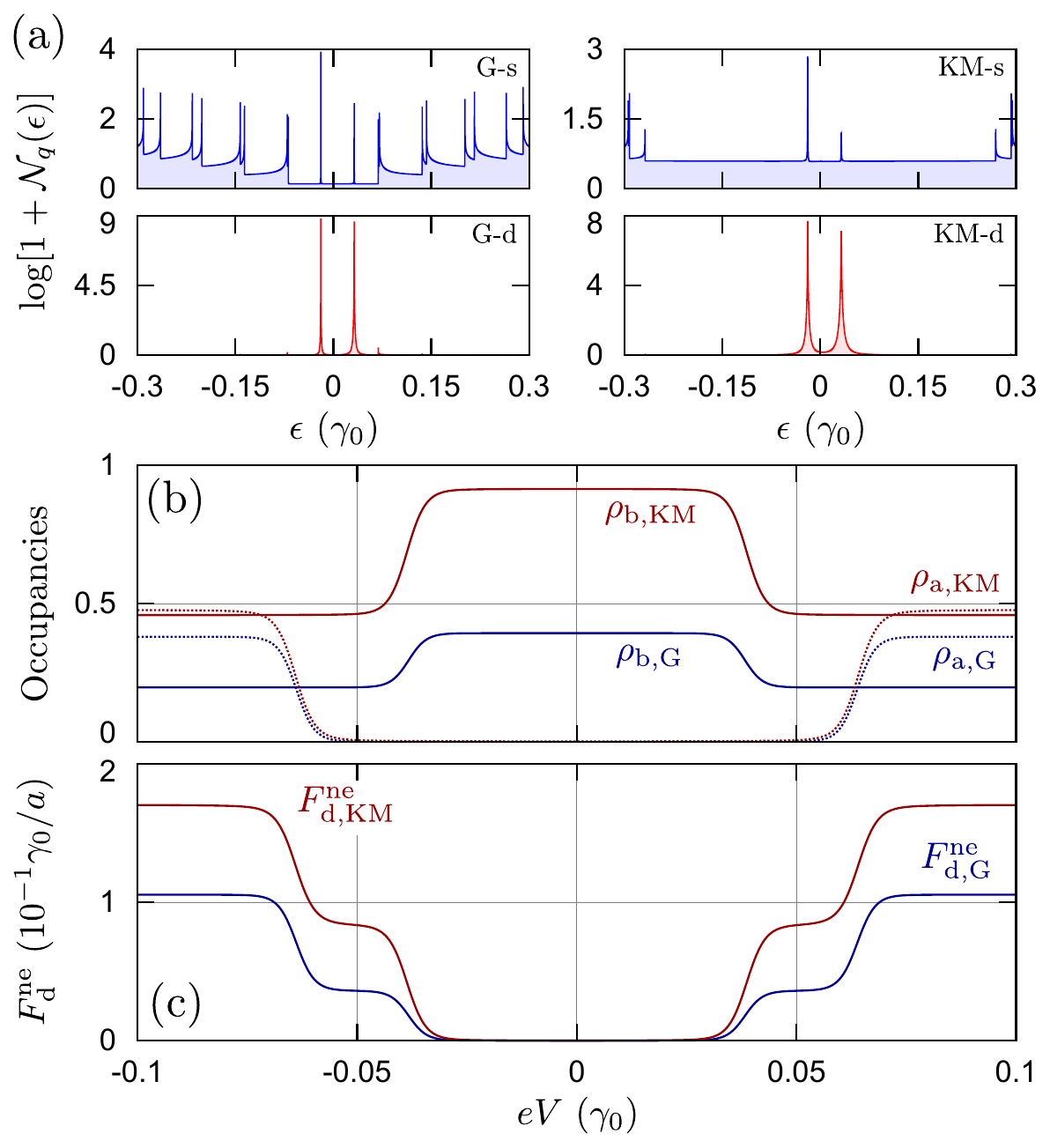}
\caption{Molecular destabilization under non-equilibrium conditions. 
(a) Density of states (DOS), using the logarithmic scale $\log(1+\mathcal{N}_q)$, projected on the substrate region around the molecule ($q=\tf{s}$, blue) and on the molecule itself ($q=\tf{d}$, red) for graphene (G, left panels) and Kane-Mele (KM, right panels) substrates. (b) Occupation probabilities of the bonding (solid lines) and antibonding (dotted lines) levels as functions of the bias voltage for graphene (G, blue lines) and Kane-Mele model (KM, red lines) substrates. (c) Corresponding non-equilibrium intramolecular force, according to Eq.~\eqref{eq:fce-d}. We use a relaxation energy $\eta = 10^{-5} \gamma_0$, an equilibrium chemical potential $\mu_0 = 0$, and $k_\tf{B} T = 10^{-3} \, \gamma_0$.}
\label{fig:rho_fza}
\end{figure}

The bond decay factor is set at $b=1.5$, allowing interaction within a radius of approximately 4.36 {\AA}, corresponding to 1\% of the maximum interaction strength, while the coupling intensity $\gamma_\tf{d}$ is tuned through $\alpha_\tf{d}=0.15$ so that the molecular levels are near the Dirac point. We select $\alpha_\tf{x}=0.5$ such that the eigenstates of the isolated molecule do not change significantly when hybridized with the substrate. In particular, for the case of graphene, this setup leads to hybridization with a single energy band near the Dirac point, composed of extended bulk states. Even though we use a larger value of $\alpha_\tf{x}$ compared to $\alpha_\tf{d}$, the qualitative results remain unchanged. As we shall discuss, this choice only affects the timescale over which the occupation dynamics takes place, without altering the underlying physical behavior. For the Kane-Mele model, we use the same spatial and coupling parameters but we add second-neighbor hoppings with $\gamma_1=0.05 \, \gamma_0$ and phase $\phi=\pm\pi/2$ to work within a topological regime. Under these conditions, the molecular energy levels fall within the topological gap and can only hybridize with localized helical states.

With these parameters, in our system the bonding level lies below the Dirac point ($\epsilon = 0$), and the antibonding level above it. If the molecule were isolated, both levels would be symmetrically placed with respect to the Dirac point. However, due to hybridization with the substrate, this balance is broken and the bonding level shifts more than the antibonding one, ending up closer to the Dirac point. This can be seen in Fig.~\ref{fig:rho_fza}(a), where we show the projected density of states (DOS), calculated as:
\begin{equation}
\mathcal{N}_q(\epsilon) = \sum_\alpha \sum_{i \in q} [\bm{\mathcal{K}}^\rho_\alpha]_{ii}(\epsilon),
\end{equation}
where $q = \{\tf{s}, \tf{d}\}$ denotes the projection region. We use the logarithmic scale $\log(1+\mathcal{N}_q)$ to reduce the high contrast produced by the bonding and antibonding peaks. The s region includes the 36 closest sites surrounding the molecule, while the d region consists only of the two molecular sites. For the d-projected DOS (on the molecule), we observe the shifted bonding peak below $\epsilon = 0$ and the antibonding peak above it. Due to the molecular hybridization with the substrate, these peaks also appear in the s-projected DOS, alongside the typical van Hove singularities of the ribbon. Regarding the magnitude, the d-projected DOS peaks are about 135 times larger than the s-projected DOS for both graphene and Kane-Mele substrates.

The large difference between the molecular and substrate densities determines a perturbative regime for the effective interaction between these subsystems. Under non-equilibrium conditions, this implies a small $F_\tf{s}$ compared with the intramolecular force.

In addition, within the framework of Refs.~\cite{brandbyge2003,leitherer2019,papior2022}, we estimate that Landauer resistivity dipoles~\cite{landauer1957}, related with the current-induced charge redistribution around the molecule, can be disregarded in a first approximation due to the observed unbalance in the projected DOS.

In Fig.~\ref{fig:rho_fza}(b) we show the occupation probability of the molecular levels as a function of the applied bias for both trivial graphene ($\gamma_1 = 0$) and topological ($\gamma_1 \neq 0$) substrates. At equilibrium ($V = 0$), the bonding level shows its maximum occupancy while the antibonding level is unoccupied. This is because we set the equilibrium chemical potential at the Dirac point, i.e., $\mu_0=0$, and the bonding (antibonding) level lies below (above) this energy point. When a bias voltage is applied and increased, the occupancy of the bonding level starts to decrease while the antibonding level becomes progressively populated. In this case, the molecular levels enter the region known as the ``bias window''---that is, between the two chemical potentials. Here, a finite electron current flows through the molecule, thereby modifying the previous equilibrium occupancies.

As can be seen, the variations in the bonding and antibonding occupancies take place at different voltage values: the bonding level depopulates first, while the antibonding occupancy begins to increase at slightly higher voltages. This difference arises from the asymmetric energy alignment of the bonding and antibonding levels with respect to $\mu_0$, as previously discussed. Importantly, both the depopulation of the bonding level and the subsequent population of the antibonding level contribute to the destabilization of the molecule. In particular, for large biases the antibonding occupancy can even surpass that of the bonding level. These results indicate that the interaction with either graphene or Kane-Mele substrates can drive molecular dissociation through electron transport. 

To account for additional relaxation processes due to the environment, we introduce a small imaginary part $\eta$ in the energy variable, i.e., $\epsilon \rightarrow \epsilon + i \eta$. This would correspond, e.g., to coupling the system to a phonon bath that allows electrons to decay into lower energy levels. The value of $\eta$ effectively sets a timescale for the electronic relaxation: a smaller $\eta$ corresponds to a longer lifetime of electronic states, while a larger $\eta$ implies faster relaxation. This allows us to capture the effect of finite-lifetime processes, such as electron-phonon interactions, without the need to model them explicitly (see Appendix~\ref{app:wb}). As a consequence of this relaxation, the molecular occupancies are not necessarily normalized. These deviate from their ideal values because we impose a finite relaxation time through the $\eta$ parameter which, in turn, reflects the decay rates to the contacts. For example, the kernel $[\bm{\mathcal{K}}_\tf{L}^\rho]_{\tf{b},\tf{b}}$, related to the bonding level's occupation from the left contact, develops a Lorentzian peak shape due to the molecule-substrate hybridization. Its width is characterized by the sum of the rate $\Gamma_{\tf{L},\tf{b}}$, associated with the molecular hybridization with the left contact, and $\eta$. Energy integration of this kernel [cf., Eq.~\eqref{eq:rho-t}] yields a factor $\Gamma_{\tf{L},\tf{b}}/(\Gamma_{\tf{L},\tf{b}}+\eta)$ which bounds the level occupancy below its ideal value. Therefore, the $\eta$ parameter establishes an additional decay mechanism that \textit{competes} with that related with the coupling to the contacts. Indeed, if we set $\eta = 0$, the level occupancy recovers its ideal value, regardless of the value of $\Gamma_{\tf{L},\tf{b}}$.

Indeed, this relaxation model could be improved. For example, in our tight-binding description of the composite system, we could implement a decoherence time or length using the D'Amato and Pastawski model~\cite{damato1990}. We estimate that such an inelastic scattering model would also yield a characteristic time scale to be compared with the molecular occupation time from the contacts.

It is often more intuitive to analyze the destabilization of the dimer in terms of the non-equilibrium force acting on the molecular bond. For this purpose we show, in Fig.~\ref{fig:rho_fza}(c), the non-equilibrium component of the intramolecular force of Eq.~\eqref{eq:fce-d}. In the absence of an applied bias, the non-equilibrium force is zero, which signals molecular stability. As the bias increases, the force exhibits a sharp rise, followed by a small shoulder over a narrow range of $V$ before rising again and reaching its maximum value at higher biases. Once more, this non-monotonic behavior, particularly the presence of the intermediate plateau, can be attributed to the shifts of the molecular levels when coupled with the substrate. Consequently, the initial increase in force is associated with the depletion of the bonding occupancy, while the subsequent rise is driven by the occupation of the antibonding level.

It may be useful to establish a comparison between the equilibrium and non-equilibrium components of the force when the two molecular levels lie within the bias window. By relating the intramolecular force with the molecular occupancies through (see Appendix~\ref{app:wb}):\footnote{Notice that for the considered parameters, the bonding and antibonding occupancies can be used here, provided the small mixing between the symmetric and antisymmetric states due to the substrate.}
\begin{equation}
F_\tf{d} \approx 2\frac{\tf{d}\gamma_\tf{d}}{\tf{d}R}(\rho_\tf{b}-\rho_\tf{a}),
\end{equation}
we obtain, for $eV>2\gamma_\tf{d}$, that $|F^\tf{ne}/F^\tf{eq}| \approx (1+\rho_\tf{a}/\rho_\tf{b})/2$. For the considered substrates we find that $\rho_\tf{a}>\rho_\tf{b}$, which implies that the destabilizing non-equilibrium component is larger than the stabilizing equilibrium term. In addition, it can be seen that $F_{\tf{d},\tf{KM}}^\tf{ne} > F_{\tf{d},\tf{G}}^\tf{ne}$, indicating a higher dissociation efficiency in the topological substrate.

It is worth noting that, in the case of the Kane-Mele substrate, the edge states exhibit spin-momentum locking, meaning that the spin polarization is tied to the propagation direction. As a result, for a given sign of the bias voltage, electrons traveling along the upper and lower edges of the ribbon carry opposite spin projections. Since the molecule is located in the vicinity of one of the edges of the ribbon, the induced destabilization is mediated by electrons with a well-defined spin polarization. Specifically, for the Kane-Mele substrate the occupation of molecular levels from a given contact (e.g., the left one) only involves spin-up electrons. In contrast, trivial graphene substrates present no such spin restriction on molecular occupation. As the number of transport channels in graphene doubles that of the Kane-Mele model, one might initially expect larger molecular occupancies in the former case. Taking into account that these occupancies scale with the hybridization of the molecular levels through the $\Gamma$ rates, the apparent contradictory result of larger occupancies in the topological case suggests that the localized nature of the topological edge states is crucial for the occupation of the molecular levels, a topic we discuss in more detail in the next section.

\subsection{Topological vs. extended bulk states}

In this section, we analyze the effect of the ribbon width on the molecular occupation, considering the extended nature of metallic states in graphene-based substrates and the localized nature of edge states in topological substrates. This analysis is particularly relevant in systems with coupling to an external environment---modeled via a finite broadening parameter $\eta$---that introduces escape channels that affect the molecular occupancies. A very small $\eta$ would act as a mathematical regularization which, for energy levels below the Fermi energy, yields occupancies close to its ideal value regardless of the system size. A finite $\eta$, on the other hand, allows the ribbon's spatial characteristics to play a physical role in the transport processes, since the decay rates $\Gamma_{\alpha,\tf{b/a}}$ modulate the molecular occupancies. This effect is illustrated in Appendix~\ref{app:fit}.

\begin{figure}[htbp]
\includegraphics[width=\columnwidth]{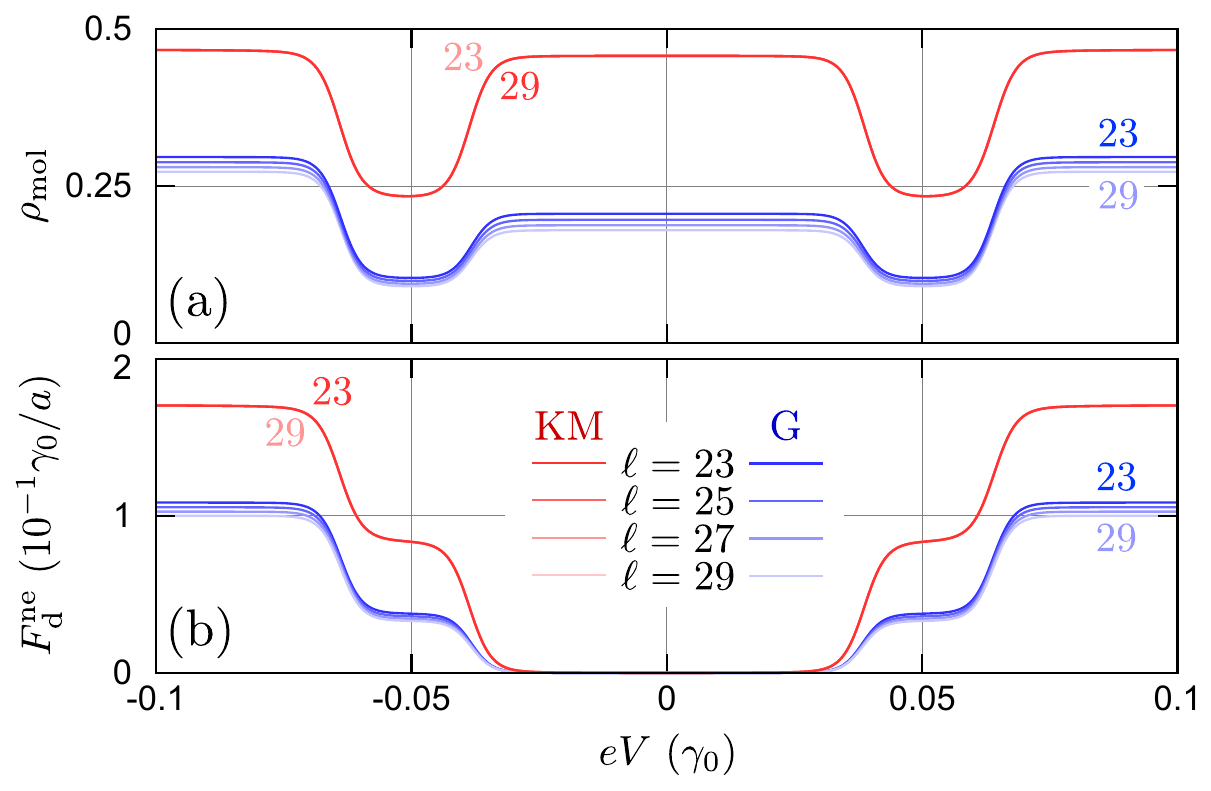}
\caption{(a) Molecular occupancy and (b) non-equilibrium intramolecular force as functions of bias voltage for varying ribbon widths, under the condition $N=3\ell+2$. Red curves represent the Kane-Mele substrate, while blue curves indicate graphene. Note, in particular, that all red curves are superimposed. The used parameters coincide with those of Fig.~\ref{fig:rho_fza}.}
\label{fig:rho_ne_tam_sustrato}
\end{figure}

To analyze the effect of the ribbon width, we calculate the half-sum of the bonding and antibonding occupancies, $\rho_\tf{mol} = (\rho_\tf{b} + \rho_\tf{a})/2$, for both substrates, as shown in Fig.~\ref{fig:rho_ne_tam_sustrato}(a). For this purpose, we use different $\ell$ values for the width of the ribbon, maintaining the metallic condition in graphene, i.e., $N = 3\ell+2$, with $\ell$ an odd number to preserve inversion symmetry with respect to the center of the unit cell. In all curves, we observe the overall behavior described in the previous section, such that the dips around $eV=\pm 0.05 \, \gamma_0$ are related with the depopulation of the bonding level and the subsequent occupation of the antibonding level, once reached by the bias window. For graphene, as the ribbon width increases, the magnitude of $\rho_\tf{mol}$ decreases, suggesting a reduced occupancy of the molecule. The observed behavior can be attributed to the normalized nature of graphene's extended states. As the ribbon width increases, the wavefunction spreads across a larger area, which results in a lower electron probability at the molecule-substrate interaction region, thus reducing the molecular occupancy. In contrast, in the topological counterpart, the edge states are inherently localized, meaning that the electron probability remains unaffected by the ribbon width. As a result, the molecular occupancy remains constant regardless of the ribbon's width.

The same behavior is obtained for the non-equilibrium intramolecular force, as we show in Fig.~\ref{fig:rho_ne_tam_sustrato}(b). The reduction in the magnitude of $F^\tf{ne}_\tf{d}$ with increasing $N$ on graphene substrates suggests a potential compromise of the material's catalytic capacity if other forces are included in the model.

The above comparison is framed in the case where the molecular states lie within a single band for graphene, and in the topological gap for the Kane-Mele model. However, when increasing the width of the nanoribbon---including the size of the contacts---additional transport channels may become available in graphene for the considered energy range. As a result, the occupancies of the molecular states can exhibit step-like increases each time a new channel becomes available. In any case, if the size of the contacts is experimentally held constant, then the surface-to-bulk ratio becomes crucial for the efficiency of the non-equilibrium catalytic process on topologically trivial substrates.

This distinction between extended and localized states can be generalized to three dimensional systems. In a 3D TI, electron transport is dominated by robust states localized at the surface, where the molecular adsorption takes place. In contrast, though a 3D trivial metal presents a continuum of bulk states that contribute to the total current, these are spread along the whole system, thereby reducing its contribution at the surface.

\subsection{Robustness against vacancy disorder}
\label{sec:disorder}

\begin{figure*}[htbp]
\includegraphics[width=\textwidth]{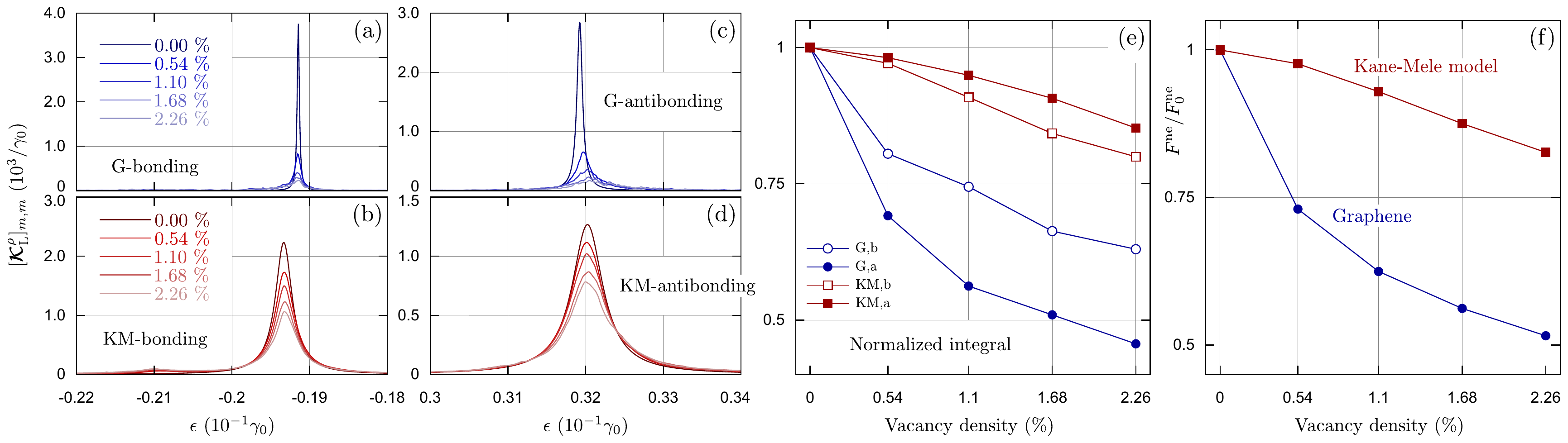}
\caption{(a)-(d) Averaged matrix element of the occupation kernel $[\bm{\mathcal{K}}^\rho_\tf{L}]_{m,m}$, indicating bonding (a,b), and antibonding (c,d) level's occupation from the left contact, for disordered graphene (a,c) and Kane-Mele (b,d) models. (e) Normalized energy integrals of the previous matrix elements. Here, blue and red lines indicate graphene and Kane-Mele cases, respectively. We use open and filled points to distinguish between bonding and antibonding occupancies. (f) Normalized non-equilibrium intramolecular force, calculated from Eq.~\eqref{eq:fce-d}, for the shown data of panel e. The parameters used coincide with those of Fig.~\ref{fig:rho_fza}, with a central region of 31 unit cells (along the transport direction) each having a width of $W=\sqrt{3}(N-1)a/2$, where $N=3 \times 25+2$.}
\label{fig:vacancies}
\end{figure*}

An exciting property of catalytic topological substrates lies in the robustness of the surface (or edge) states against different types of disorder including, e.g., those associated with catalyst poisoning. To explore this feature, we study the effect of substrate degradation on catalytic performance, analyzing how the presence of atomic vacancies in both trivial (graphene) and topological (Kane-Mele model) substrates impacts molecular stability. In order to do so, we incorporate random carbon vacancies along the central system, while keeping pristine contacts [see Fig.~\ref{fig:scheme}(a)]. In our simulations, to stabilize the excess charge introduced by the vacancies, we passivate the nanoribbon by removing any carbon atoms that remain connected to the lattice through a single (nearest neighbor) hopping element.

When introducing carbon vacancies, the standard procedure \cite{he2013,lee2014} is to remove both the onsite and all hopping terms connected to the corresponding site (including nearest- and next-nearest-neighbor hoppings). In doing so, the phase from the next-nearest-neighbor hoppings is  symmetrically removed from both spin subspaces, thus preserving time-reversal symmetry.

To establish a representative number of vacancies in the substrate, we first fix a ``target'' vacancy rate (e.g., 2 \%) and generate a large number of samples by removing lattice sites with that probability. In each configuration, we subsequently remove dangling bonds, which slightly increases the total number of vacancies. We then count the final number of vacancies in each configuration after this cleaning process, to obtain the most probable number of vacancies for a given target rate. We then filter the ensemble of configurations and retain only those with exactly that number of vacancies. This procedure allows us to average physical quantities over a statistically meaningful set of samples with the \textit{same} final vacancy count, ensuring consistency across the ensemble.

In Figs.~\ref{fig:vacancies}(a)-(d) we show the bonding and antibonding occupation kernels $[\bm{\mathcal{K}}^\rho_\tf{L}]_{m,m}$ related to the left contact, within a reduced energy window around the molecular level, and averaged over 1000 independent disorder configurations. The computation of the average occupancies is motivated by the physical scenario in which a large number of molecules are deposited around one of the edges of the substrate, far from the contacts, in such a way that interactions between them can be neglected (i.e., low molecular concentration). Consequently, each vacancy configuration describes the substrate environment around a single molecule. In the case of graphene (panels a and c), we observe a significant reduction in the intensity of the occupation kernels, even for low vacancy concentrations such as 0.54 \%. This reduction is accompanied by a broadening of the peak, mainly due to a random shift of the molecular level around its original value for the pristine sample. These effects can be attributed to a high sensibility of the substrate wavefunction to the particular distribution of the vacancies. Even at a fixed concentration of vacancies, the energy position, height and width of the kernel peak strongly fluctuates between different samples, yielding high standard deviation values related with their averages.
In contrast, the Kane-Mele model in Figs.~\ref{fig:vacancies}(b) and (d) shows a more robust behavior: although there is a decrease in the intensity with increasing vacancy concentration, this reduction is much less pronounced compared to graphene samples. Moreover, the position and width of the average peaks remain nearly unchanged, indicating the robustness of the topological edge states to disorder.

Figure~\ref{fig:vacancies}(e) summarizes the above analysis, showing the energy-integrated occupation kernels for both bonding and antibonding levels, normalized by the corresponding values of the pristine sample. Although these integrals were taken in a reduced energy range around the kernel peaks, they still capture the main contribution to the occupancies. Therefore, their normalized values reflect the sensitivity of the molecular occupation to disorder. For the Kane-Mele substrate, we observe an almost linear reduction, reaching 80\% and 85\% of the pristine bonding and antibonding values at 2.26\% vacancy concentration, respectively. In contrast, for graphene, the decrease is more abrupt, showing a sublinear behavior to about 63\% and 46\% of the pristine bonding and antibonding values, respectively, at the same vacancy concentration.

The faster attenuation of the average kernel peaks in graphene samples reflects the high sensitivity of the molecular occupancies to the spatial distribution of disorder, which varies across samples even at fixed vacancy count. In contrast, the Kane-Mele model maintains a consistent spectral structure even at vacancy concentrations above 2\%, a value where many properties characteristic in graphene based ribbons are lost. This suggests that the topological nature of the edge states plays a crucial role in their response to disorder.

To further support these results, we use the previously calculated occupancies to show, in Fig.~\ref{fig:vacancies}(d), the non-equilibrium intramolecular force (normalized to its pristine value $F^\tf{ne}_0$) for a situation where the bias window includes both molecular levels. Here, we observe that in graphene substrates this force is clearly attenuated as the vacancy concentration increases, reaching 50\% of its pristine value at 2.26\% concentration. For topological substrates, however, the destabilizing force only dropped to about 80\% at the same vacancy count, thus reinforcing the previously observed stability of the molecular occupancies. These findings highlight the robustness of the considered topological substrate as a catalyst, even at high vacancy concentrations, contrasting with the higher sensitivity to disorder observed in trivial graphene.

\section{Conclusions}

In this work, we investigated the interaction between a molecular dimer and graphene nanoribbons used as substrates under non-equilibrium conditions. By considering spin-orbit coupling within the substrate, we explored molecular dissociation in the context of the emerging non-trivial topology. For this purpose, we calculated the occupancies of the bonding and antibonding molecular levels hybridized with the surrounding substrate sites, and related these occupancies to the intramolecular force acting between its constituent atoms.

From a transport perspective, we identified a key mechanism associated with the destabilization of the molecule. At zero bias voltage, the system is in a stable equilibrium state where the bonding level sits below the Fermi energy. Increasing the bias window then enables the depopulation of the bonding level and the subsequent occupation of the antibonding level, which opposes the equilibrium attractive force of the molecular bond. These occupancies have different decay rates due to the specific hybridization of the molecular levels with the substrate states. Interestingly, we found that in both trivial (graphene without SOC) and topological (with SOC) substrates, the antibonding occupancy exceeds that of the bonding level. This implies, in particular, that the magnitude of the non-equilibrium force exceeds that of the equilibrium force when the bias window includes both molecular levels. In the more general frame of multi-electronic states, these hybridized molecular levels can be associated with frontier molecular orbitals (HOMO and LUMO). In this context, the transport processes can be interpreted as intramolecular redox reactions, where the HOMO oxidizes to enable the reduction of the LUMO.

In the above scenario, we contrasted the role of the bulk (extended) states in graphene with the topological (localized) edge states in the Kane-Mele model, focusing on their effect on molecular occupancies and the related intramolecular force. We observe that while an increase in the ribbon's width does not affect the molecular occupancies in topological substrates, in trivial graphene, both the bonding and antibonding occupancies are reduced, thus compromising the effectiveness of the catalytic reaction. This is attributed to the normalized electron probability across the unit cell. Since the molecule only interacts with a small region near one of the edges of the substrate, the normalization of the extended states reduces the molecular occupancies as the ribbon's width increases. In this sense, we find a reduction of the catalytic efficiency in non-topological substrates, related to the ratio between the active surface and the volume of the transport region.

Finally, we investigated the molecular stability against substrate disorder by introducing random vacancies in the central region of the transport device. In line with the expected topological robustness of the localized edge states, our simulations reveal a sustained behavior of the molecular occupancies for the Kane-Mele substrate, even as vacancy concentration increases. In contrast, graphene demonstrates a high sensitivity to disorder. We found qualitative differences in both the normalized occupancies and the non-equilibrium intramolecular force. The Kane-Mele model shows a slow, nearly linear decay, whereas the trend in graphene is more pronounced and sublinear. Our findings reveal that the catalytic capacity of the topological model is greater than that of trivial graphene for pristine samples, and the inclusion of disorder further enhances this effect.

A detailed study of an ensemble of molecular configurations lies beyond the scope of the present work. Nevertheless, as we discuss in Appendix \ref{app:angle}, for the topological substrate the average behavior would approximately correspond to the individual molecular response. In contrast, for a non-topological substrate, this behavior may not be trivially predictable, since even at low concentrations the ensemble could act as an additional source of disorder.

In summary, we found qualitative signatures related to the topological nature of the substrate, which can be extended to three-dimensional systems. In this sense, the minimal model proposed in this work allows for a simplified description of the physics behind non-equilibrium heterogeneous topological catalysis.

\section{Acknowledgements}

M.B. acknowledges financial support from Consejo Nacional de Investigaciones Cient\'{i}ficas y T\'{e}cnicas (CONICET, PIBA 28720210100973CO), and Secretar\'{i}a de Ciencia y Tecnolog\'{i}a (SECyT-UNC) through Grant No. 33820230100101CB. E.L.M. and H.L.C. acknowledge financial support from CONICET (PIP 2022-59241), and SECyT-UNC through Grant No. 33620230100363CB.

\appendix

\section{Occupation mechanisms in the wideband limit}
\label{app:wb}

\subsection{Explicit calculations}

\begin{figure}[!ht]
\includegraphics[width=0.95\linewidth]{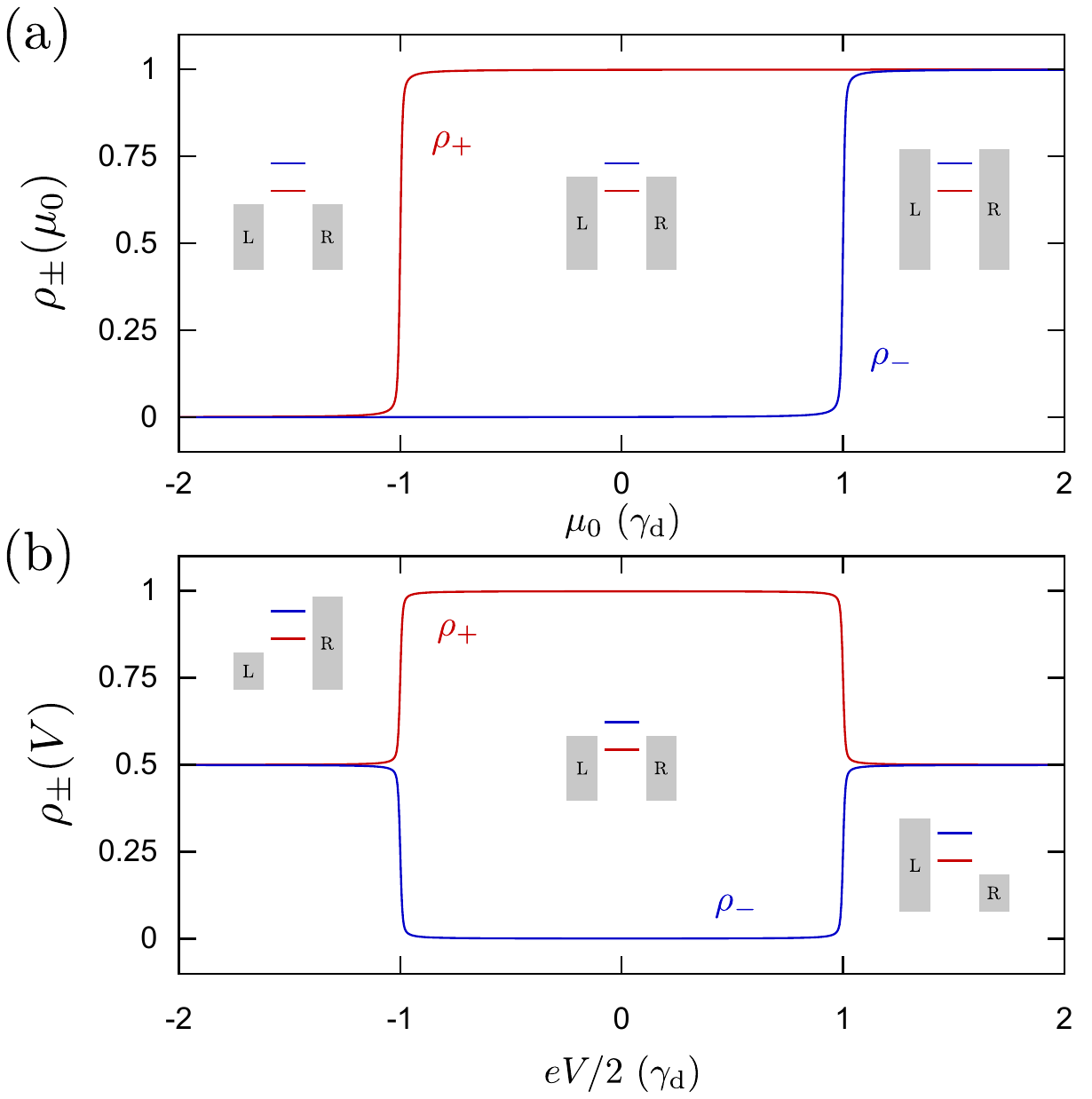}
\caption{Bonding (red) and antibonding (blue) occupancies under (a) equilibrium and (b) non-equilibrium conditions. The used parameters are: $E_\tf{d} = 0$, $\Gamma = k_\tf{B} T = 10^{-3} \, \gamma_0$, $\gamma_\tf{d} = 0.3 \, \gamma_0$, and $\eta = 0$, with $\gamma_0$ a reference energy. The schemes denote the contacts chemical potentials and the molecule's energy levels.}
\label{fig:eta0}
\end{figure}

We consider a diatomic molecule symmetrically coupled to left (L) and right (R) reservoirs in the wideband limit. In the local site basis $\{\ket{1}, \ket{2}\}$, the effective Hamiltonian can be written as:
\begin{equation}
\bm{H}_\tf{eff} =
\begin{pmatrix}
E_\tf{d} - i\Gamma & -\gamma_\tf{d} \\
-\gamma_\tf{d} & E_\tf{d} - i\Gamma
\end{pmatrix},
\label{eq:ham}
\end{equation}
where $E_\tf{d}$ denotes the onsite energy of each atomic orbital, $\gamma_\tf{d}$ is the hopping amplitude between the two atoms, and $\Gamma$ represents the coupling strength to the leads, assumed to be equal at both sites and energy-independent due to the wideband approximation. The bonding ($+$) and antibonding ($-$) eigenenergies and eigenstates are given by:
\begin{equation}
E_\pm = E_\tf{d} \mp \gamma_\tf{d}, \quad \ket{\pm} = \frac{\ket{1} \pm \ket{2}}{\sqrt{2}}.
\label{eq:ba-basis}
\end{equation}
In this diagonal basis $\{\ket{+},\ket{-}\}$, the retarded Green's function writes $\bm{G}^r(\epsilon) = \tf{diag}[1/(\epsilon - E_+ + i\Gamma),1/(\epsilon - E_- + i\Gamma)]$, while the coupling matrices to the contacts take the form:
\begin{equation}
\bm{\Gamma}_\tf{L} = \frac{\Gamma}{2}
\begin{pmatrix}
1 & 1 \\
1 & 1
\end{pmatrix}, \quad
\bm{\Gamma}_\tf{R} = \frac{\Gamma}{2}
\begin{pmatrix}
1 & -1 \\
-1 & 1
\end{pmatrix}.
\end{equation}
The bonding and antibonding occupancies can thus be obtained from Eq.~\eqref{eq:rho-t}, where we have:
\begin{equation}
\rho_\pm = \int_{-\infty}^{\infty} \frac{\tf{d}\epsilon}{\pi} \frac{\Gamma}{(\epsilon - E_\pm)^2 + \Gamma^2} \bar{f}(\epsilon),
\end{equation}
with $\bar{f}=(f_\tf{L}+f_\tf{R})/2$. In Fig.~\ref{fig:eta0}(a), we show the equilibrium occupancies as functions of the chemical potential $\mu_0$ for an intermediate coupling regime. As $\mu_0$ increases, the occupancies begin to rise once the chemical potential exceeds the energy of each molecular state, reaching a maximum value of 1 and then remaining constant. Each of these regimes is illustrated schematically in the figure.

Fixing $\mu_0 = 0$, we present in Fig.~\ref{fig:eta0}(b) the molecular occupancies under a finite bias $V = (\mu_\tf{L}-\mu_\tf{R})/e$ applied across the system. It can be seen that, for $|eV|/2 < \gamma_\tf{d}$, the contribution to the antibonding state is practically zero, while the bonding state remains nearly fully occupied. When the applied voltage exceeds this threshold, a finite current is established through the system, and the occupancies of both states become equal to 1/2. Importantly, the magnitude of these occupancies does not depend on the direction of the current induced by the bias. Assuming a positive bias, the left contact is able to populate both molecular states equally, while the chemical potential of the right contact lies below the energy of both levels. As a result, electrons occupying these states can decay into the right contact, leading to an occupancy of 1/2 for each state.

In general systems, the above discussion can be extended to better understand the maximum value of the occupancies. By assuming the equilibrium condition, such that $\bar{f} = f_0$, the density matrix can be written in terms of the spectral function $\bm{A}=i (\bm{G}^r-\bm{G}^a)$ as:
\begin{equation}
\bm{\rho} = \int_{-\infty}^\infty \frac{\tf{d}\epsilon}{2\pi} \bm{A}(\epsilon) f_0(\epsilon),
\end{equation}
which, for the diagonal elements, yields:
\begin{equation}
\rho_{ii} = \int_{-\infty}^\infty \tf{d}\epsilon \, \mathcal{N}_i(\epsilon) f_0(\epsilon).
\end{equation}
with $\mathcal{N}_i = -\tf{lim}_{\eta \to 0^+}\tf{Im}[G^r_{ii}(\epsilon+i\eta)]/\pi$ the local density of states projected on site $i$. By using the spectral decomposition of the Green's function, we find
\begin{equation}
\mathcal{N}_i(\epsilon) = -\lim_{\eta \to 0^+} \sum_k |\!\braket{i|k}\!|^2 \frac{\eta}{(\epsilon - \epsilon_k)^2 + \eta^2}.
\end{equation}
This expression describes a sum over Lorentzian functions centered at the eigenenergies $\epsilon_k$. 
In the limit where the chemical potential lies well above all the system's energy levels ($\mu_0\gg\max[\epsilon_k]$), the Fermi function can be approximated as $f_0(\epsilon) \approx 1$, yielding
\begin{equation}
\rho_{ii} = \sum_k |\!\braket{i|k}\!|^2 \lim_{\eta \to 0^+} \int_{-\infty}^{\infty} \frac{\tf{d}\epsilon}{\pi} \frac{\eta}{(\epsilon-\epsilon_k)^2 + \eta^2}=1.
\label{eq:mer}
\end{equation}
It is important to note that this result relies on the assumption that the set of eigenstates $\{ \ket{k} \}$ forms a complete basis for the local site $\ket{i}$, and that all these states are accessible through the leads. However, this assumption may fail. For instance, if the system hosts localized states that are disconnected from the leads, the spectral function can reflect this by suppressing the weight of such states at the contact site. Physically, this means that an electron injected from the lead cannot reach the localized state, and the corresponding contribution to the occupation is lost. In terms of the spectral decomposition, the sum over $|\!\braket{i|k}\!|^2$ in Eq.~\eqref{eq:mer} may then be less than one. As a result, the integral of the local density of states over all energies can yield a value strictly below one. This effect is particularly relevant in systems with disordered substrates, where localization due to lattice defects may appear \cite{pereira2006}.

\subsection{Finite $\eta$-broadening effects}

Under certain conditions, the occupancies of the molecular levels do not necessarily reach the ideal value of 1, even when these fall well below the  chemical potentials $\mu_\alpha$ and in the absence of localization effects. This is due to a competing effect with additional decay channels not included in the $\alpha$ sum of Eq.~\eqref{eq:rho-t}. In our simulations, it is common to introduce a small imaginary term $\eta$ in the energy variable to regularize integrals: $\epsilon \to \epsilon + i\eta$. This introduces a level broadening that adds to the physical $\Gamma$. To preserve the ideal occupancies the numerical regulator $\eta$ must be much smaller than $\Gamma$. If $\eta$ is not small enough, it introduces a typical timescale for the electron processes which has to be compared with $\Gamma$.

\begin{figure}[!ht]
\includegraphics[width=0.95\linewidth]{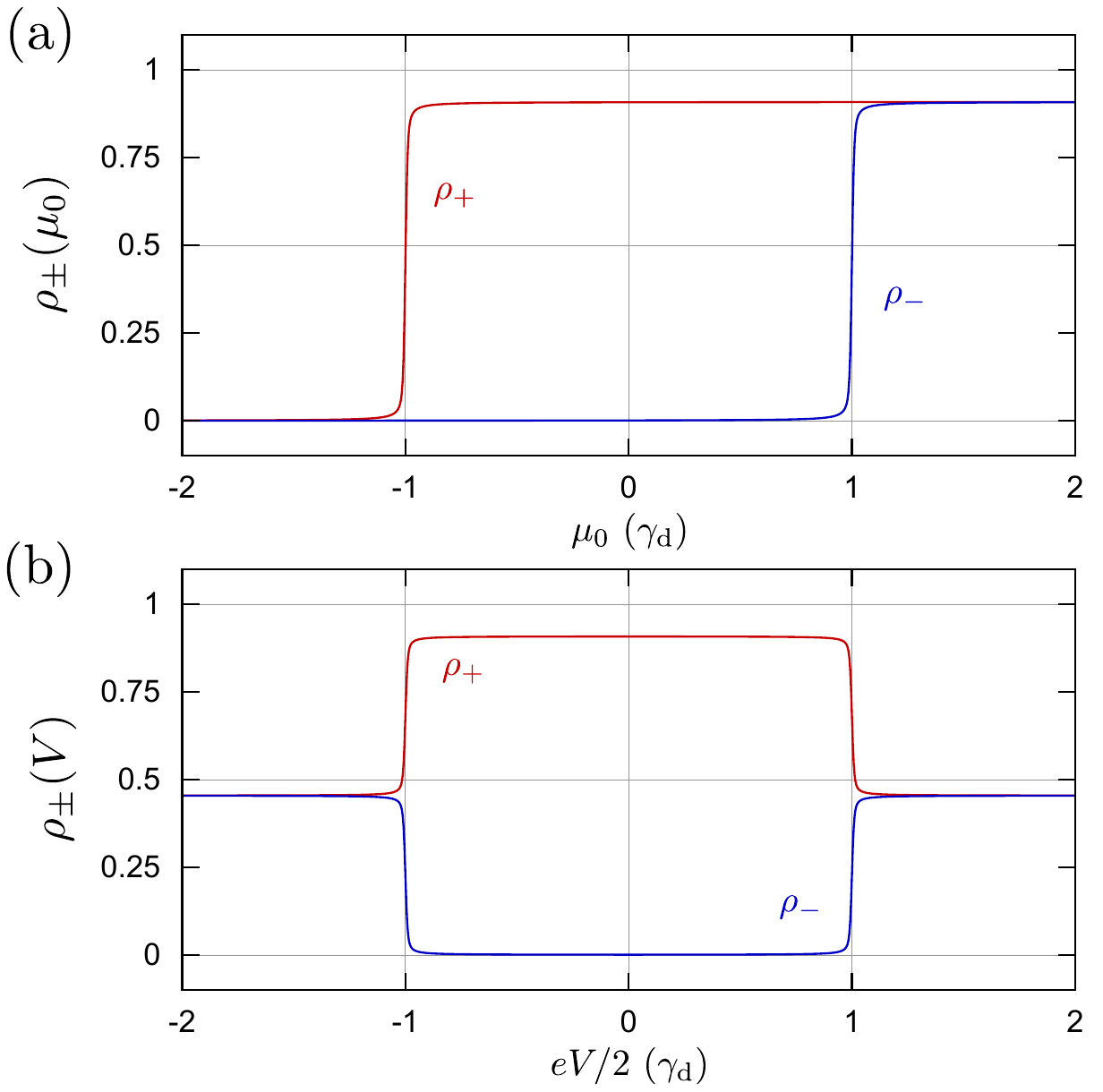}
\caption{Bonding (red) and antibonding (blue) occupancies under (a) equilibrium and (b) non-equilibrium conditions for a finite $\eta$ parameter. The used parameters coincide with those of Fig.~\ref{fig:eta0}, though we use $\eta = 10^{-4} \, \gamma_0$.}
\label{fig:etafin}
\end{figure}

A finite $\eta$ can be interpreted as an effective coupling to an external environment acting as a thermal bath, which introduces energy relaxation in the system. The broadening introduced by $\eta$ thus represents a finite lifetime of the electronic states due to interactions with the additional environment, which is not taken into account in the $\alpha$ sum of Eq.~\eqref{eq:rho-t}. In consequence, the occupancies of the bonding and antibonding states are:
\begin{equation}
\rho_\pm = \int_{-\infty}^{\infty} \frac{\tf{d}\epsilon}{\pi} \frac{\Gamma}{(\epsilon - E_\pm)^2 + (\Gamma+\eta)^2} \bar{f}(\epsilon).
\end{equation}
This modifies the Lorentzian kernel in the occupation integrals, leading to a reduced amplitude. Specifically, since the numerator remains proportional to $\Gamma$ while the denominator grows as $(\Gamma + \eta)^2$, the overall weight of the peak decreases as $\Gamma/(\Gamma+\eta)$. This results in lower occupancies for the bonding and antibonding states, as shown in Fig.~\ref{fig:etafin}. Physically, this reduction in the occupancies can be understood as a consequence of energy relaxation caused by the environment. An electron injected from a contact has a finite probability of being absorbed or scattered by the bath before fully occupying a molecular state.

Interestingly, if the environment were instead included as an additional source, defined with a corresponding distribution $f_\tf{bath}(\epsilon)$, the total occupancy would again recover its maximum value. However, since $\eta$ typically models an incoherent, uncontrolled environment, this reflects an irreversible leakage of electronic population into external degrees of freedom.

\subsection{Electronic force}

For the computation of the force, we consider $\gamma_\tf{d}(x)>0$ from Eq.~\eqref{eq:ham} to be a decreasing function of the bond length $x=2R$, see Eq.~\eqref{eq:hop_dim}. The force operator, defined as the derivative of the Hamiltonian with respect to the bond length, takes the explicit form in the site basis $\{\ket{1},\ket{2}\}$:
\begin{equation}
\bm{\Lambda}=-\frac{\partial\bm{H}}{\partial x} = \frac{\tf{d}\gamma_\tf{d}}{\tf{d}x}
\begin{pmatrix}
0 & 1 \\
1 & 0
\end{pmatrix}.
\end{equation}
On the other hand, from Eq.~\eqref{eq:rho-t} we define the occupation kernel as $\bm{\mathcal{K}}^\rho_\alpha=\bm{G}^r \bm{\Gamma}_\alpha \bm{G}^a/\pi$, such that the force kernel reads:
\begin{equation}
\mathcal{K}^F_\alpha = \tf{tr}[\bm{\Lambda}\bm{\mathcal{K}}^\rho_\alpha]=2\frac{\tf{d}\gamma_\tf{d}}{\tf{d}x}\tf{Re}\,[\bm{\mathcal{K}}^\rho_\alpha]_{12}.
\end{equation}
By rewriting the states $\ket{1}$ and $\ket{2}$ in terms of the bonding and antibonding basis $\{\ket{+},\ket{-}\}$ [Eq.~\eqref{eq:ba-basis}], associated with the symmetric and antisymmetric combinations of atomic orbitals, we obtain:
\begin{equation}
\mathcal{K}^F_\alpha = \frac{\tf{d}\gamma_\tf{d}}{\tf{d}x} \left( [\bm{\mathcal{K}}^\rho_\alpha]_{++} - [\bm{\mathcal{K}}^\rho_\alpha]_{--} \right).
\end{equation}
With the force kernel now expressed in terms of the occupation kernels, we can use Eq.~\eqref{eq:force} to relate the force directly to the molecular occupancies:
\begin{equation}
F = \frac{\tf{d}\gamma_\tf{d}}{\tf{d}x} (\rho_{+} - \rho_{-}).
\label{eq:fce}
\end{equation}
Given that $\tf{d}\gamma_\tf{d}/\tf{d}x < 0$ and the diagonal elements of the occupation kernel are positive, this result shows that the occupancy of the bonding level contributes as an attractive force, while that of the antibonding level contributes as a repulsive force.

\begin{figure}[!ht]
\includegraphics[width=0.95\linewidth]{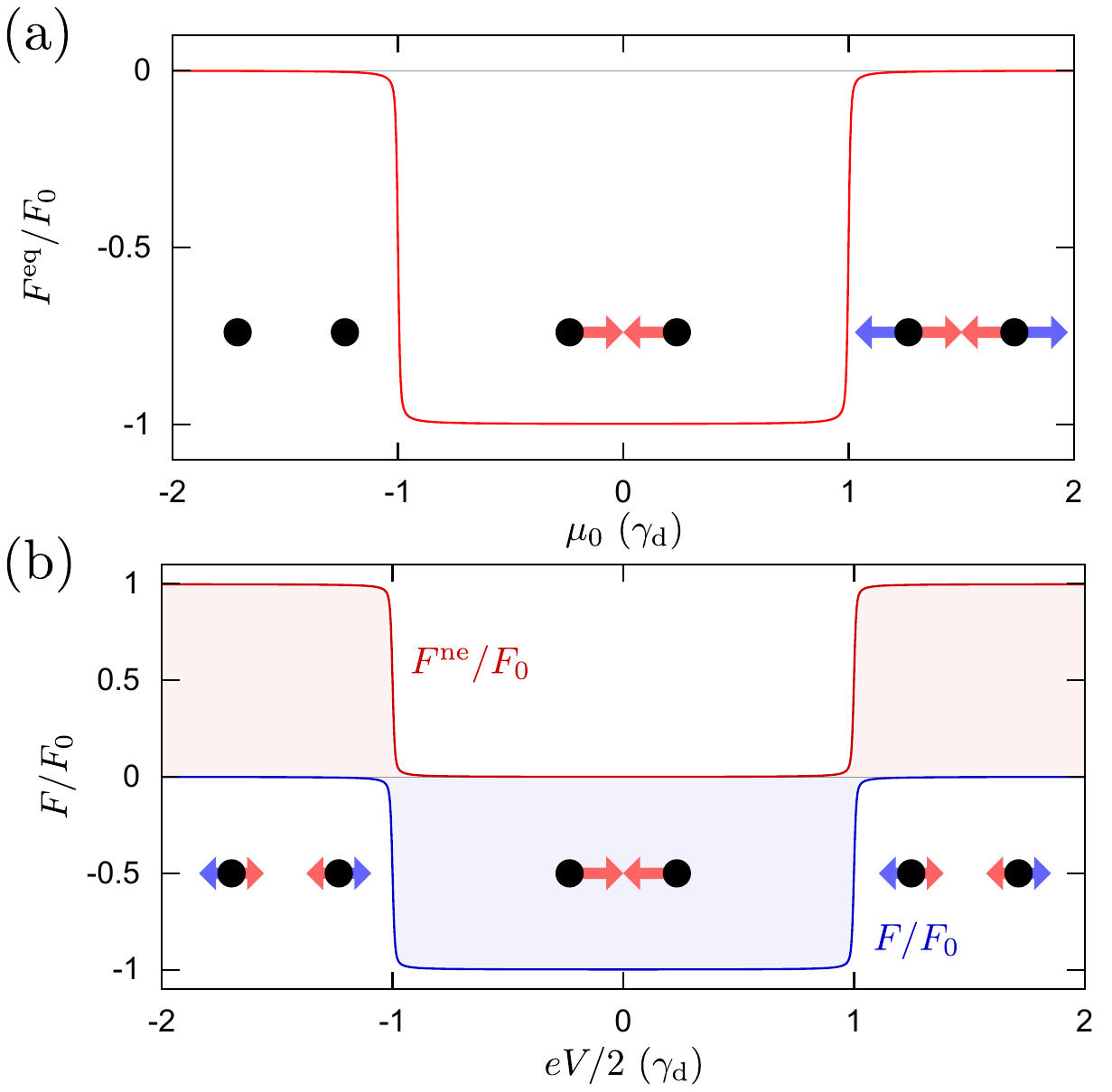}
\caption{Normalized force under (a) equilibrium and (b) non-equilibrium conditions for the situation shown in Fig.~\ref{fig:eta0}. In panel (b) we show the non-equilibrium contribution in red and the total force in blue. The schemes in both panels denote the bonding (red arrows) and antibonding (blue arrows) force contributions.}
\label{fig:fce}
\end{figure}

In Fig.~\ref{fig:fce}(a) we show the equilibrium force ($\bar{f}=f_0$) as a function of the equilibrium chemical potential. For $|\mu_0| > \gamma_\tf{d}$ the force vanishes. This occurs because the bonding and antibonding states are either both empty $\mu_0 < -\gamma_\tf{d}$ or fully occupied $\mu_0 > \gamma_\tf{d}$. In the former case, neither state contributes to the force, while in the latter, both contribute equally and oppositely, resulting in complete cancellation (see schemes in the figure). For $|\mu_0| < \gamma_\tf{d}$, the force reaches its minimum value given that in that region only the bonding state is occupied. This contribution is attractive, stabilizing the molecular bond, according to Eq.~\eqref{eq:fce}. To maintain molecular stability, this equilibrium force must be balanced by other forces---such as those between the atomic nuclei---which are not considered in our model. It is important to mention that in this simplified analysis, the simultaneous occupation of the bonding and antibonding levels is not prohibited, since our formalism is restricted to the single electron problem. In a more involved calculation, however, occupation of the molecular levels would imply an additional energy cost related with the electron Coulomb repulsion.

In Fig.~\ref{fig:fce}(b) we take $\mu_0 = 0$ and a finite bias voltage is applied. In this case, the total force is dominated by the bonding contribution in the region $|eV/2| < \gamma_\tf{d}$ and it becomes zero when $|eV/2| > \gamma_\tf{d}$, reflecting the destabilizing effect of the non-equilibrium configuration. In this latter regime, both the bonding and antibonding levels become accesible within the bias window, yielding occupancies of 1/2, as shown in Fig.~\ref{fig:eta0}, such that the initial atractive force is reduced by the depopulation of the bonding level and compensated with the repulsive term coming from the antibonding occupancy.

\textit{Graphene based substrates.--} We can extend the previous analysis by considering graphene and Kane-Mele substrates in which the same relation holds for the intramolecular force and the level occupancies. The difference here is that the molecular levels are hybridized with the substrate. In addition, we need to include different spin channels in our setup. In this case the total Hilbert space doubles, and the force kernel now reads:
\begin{equation}
\mathcal{K}^F_{\alpha,\sigma} = \frac{\tf{d}\gamma_\tf{d}}{\tf{d}x} \left( \mathcal{K}^\rho_{\alpha,\sigma,+} - \mathcal{K}^\rho_{\alpha,\sigma,-} \right),
\end{equation}
with $\sigma=\{\uparrow,\downarrow\}$ accounting for the electron's spin projection. By assuming spinless chemical potential of the contacts, i.e., $\mu_{\alpha,\sigma} = \mu_\alpha$, the intramolecular force takes the form:
\begin{equation}
F = \frac{\tf{d}\gamma_\tf{d}}{\tf{d}x} \sum_\sigma (\rho_{+,\sigma} - \rho_{-,\sigma}) = 2\frac{\tf{d}\gamma_\tf{d}}{\tf{d}x} (\rho_{+} - \rho_{-}),
\end{equation}
where we define $\rho_m=\sum_\sigma \rho_{m,\sigma}/2$, with $m=\{+,-\}$. We now analyze the non-equilibrium force in a situation where the two molecular levels fall within the bias window and compare it with the equilibrium situation where only the symmetric level falls below $\mu_0$. For $V=0$ we can approximate the equilibrium force as:
\begin{equation}
F = F^\tf{eq} \approx  \frac{\tf{d}\gamma_\tf{d}}{\tf{d}x} \sum_\sigma \bar{\rho}_{+,\sigma} \, ,
\end{equation}
where $\bar{\rho}_{m,\sigma}$ means the energy integral of the occupation kernels around the level peak at the eigenenergy $E_m$, i.e.,
\begin{equation}
\bar{\rho}_{m,\sigma} = \sum_\alpha \int  \mathcal{K}^\rho_{\alpha,\sigma,m}(\epsilon) \, \tf{d}\epsilon .
\end{equation}
On the other hand, for positive bias $eV>2\gamma_\tf{d}$ we have
\begin{equation}
F = F^\tf{eq} + F^\tf{ne} \approx  \frac{\tf{d}\gamma_\tf{d}}{\tf{d}x} \sum_\sigma \frac{\bar{\rho}_{+,\sigma} - \bar{\rho}_{-,\sigma}}{2},
\end{equation}
where the factor $1/2$ comes from the fact that here only the left contact contributes to the occupation of the molecular levels and the system preserves mirror symmetry. Therefore, by subtracting the equilibrium term of the force, we obtain for the non-equilibrium force:
\begin{equation}
F^\tf{ne} \approx  -\frac{\tf{d}\gamma_\tf{d}}{\tf{d}x} \sum_\sigma \frac{\bar{\rho}_{+,\sigma} + \bar{\rho}_{-,\sigma}}{2}.
\end{equation}
Once more, since $\tf{d}\gamma_\tf{d}/\tf{d}x$ is negative, the equilibrium force always acts to compress the molecule, while the non-equilibrium component promotes its destabilization.

\section{$\theta$-dependence of molecular occupation and force}

\label{app:angle}

\begin{figure}[!ht]
\includegraphics[width=\columnwidth]{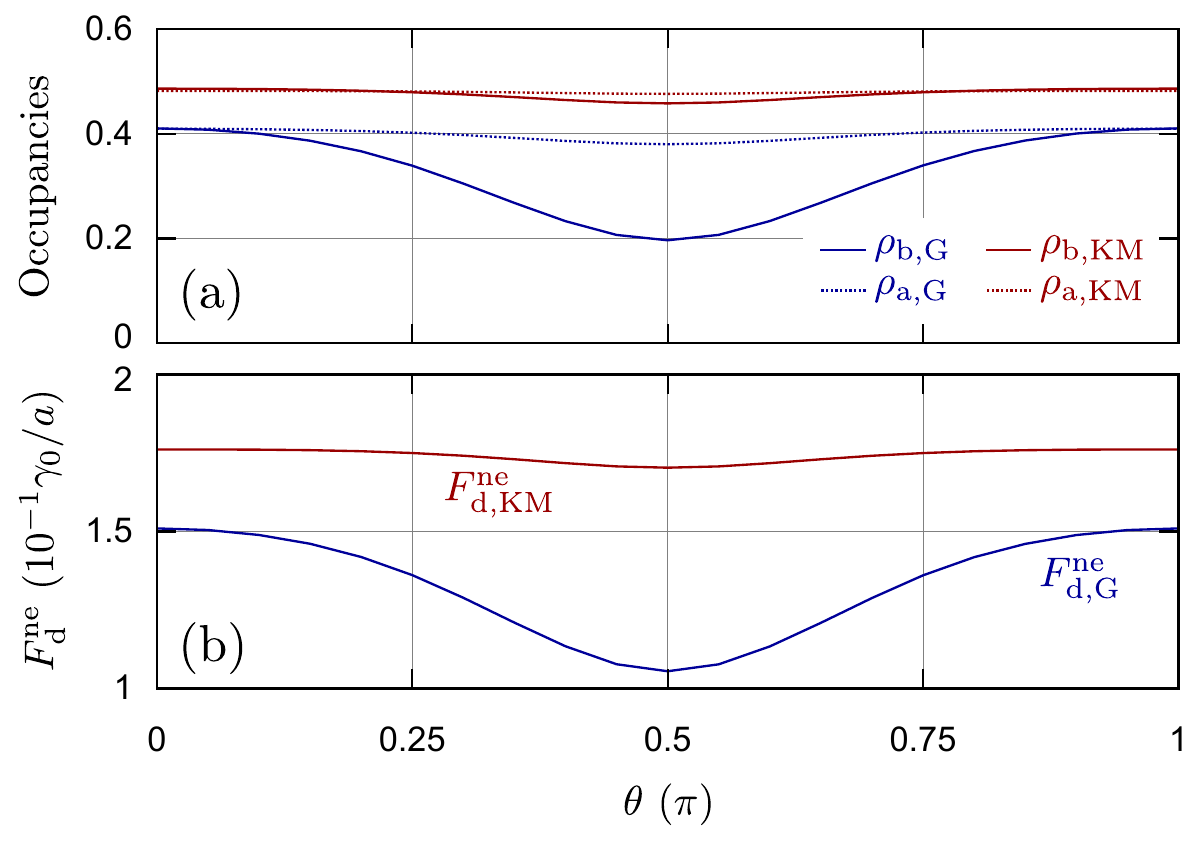}
\caption{(a) Bonding (solid lines) and antibonding (dotted lines) occupancies for graphene (blue) and Kane-Mele (red) substrates as functions of the molecular angle $\theta$ [see Fig.~\ref{fig:scheme}(a)]. (b) Corresponding non-equilibrium intramolecular forces. We fix the bias voltage at $V = 0.1 \, \gamma_0$ and keep the other parameters the same as in Fig.~\ref{fig:rho_fza}.}
\label{fig:angle}
\end{figure}

In Fig.~\ref{fig:angle}, we show the molecular occupancies (panel a) and the intramolecular force $F_\tf{d}$ (panel b) for a fixed bias as functions of the molecule's angle $\theta$ subtended between the molecular bond direction and the transport axis, according to the inset of Fig.~\ref{fig:scheme}(a) and Eq.~\eqref{eq:position}. In particular, the case $\theta = \pi/2$ corresponds to that of Figs.~\ref{fig:rho_fza}(b) and (c) for $V = 0.1 \, \gamma_0$.

As expected (since we consider a homonuclear dimer), both the molecular occupancies and the intramolecular force are symmetric functions of $\theta$ around $\pi/2$. These quantities exhibit minimal variation when the molecule is placed on a Kane-Mele substrate. Conversely, on a graphene substrate, a larger amplitude is observed for the bonding occupancy, resulting in a more pronounced related force. Nevertheless, this force remains positive (repulsive) across the full angular range, and is smaller than that of the Kane-Mele substrate.

The above discussion allows to draw an initial hypothesis on how the obtained results would change when considering an ensemble of molecules with different orientations and positions near the ribbon's edge. In the limit of low molecular concentrations (negligible intermolecular interactions), and considering the low sensitivity to disorder on the dissociative force observed for topological substrates (see Section \ref{sec:disorder}), it can be anticipated that the ensemble behavior would approximately correspond to the average of the individual molecular responses on such substrates. In contrast, for a non-topological substrate, the strong dependence of the current-induced dissociative forces on disorder suggests that the overall ensemble behavior may not be trivially predictable, since even at low concentrations, the ensemble of molecules could act as an additional source of disorder.\\

\section{Molecular occupancies and substrate size}
\label{app:fit}

\begin{figure*}[!ht]
\includegraphics[width=\textwidth]{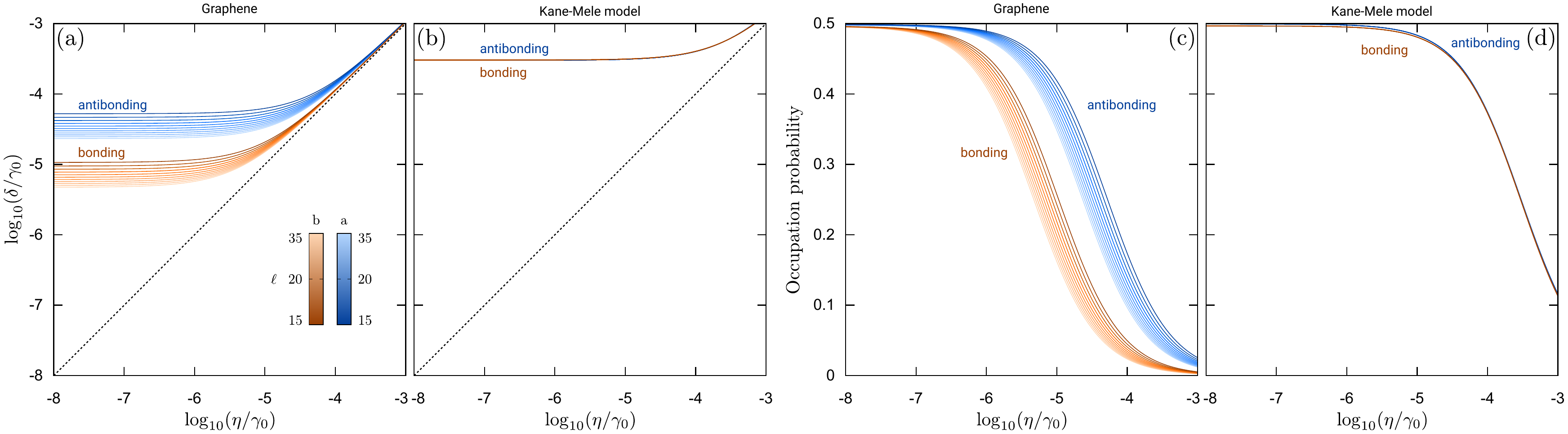}
\caption{Occupation rates for graphene (a) and Kane-Mele (b) substrates as functions of the relaxation rate $\eta$ for different ribbon widths, plotted in log-log scale. (c) and (d) show the corresponding occupancies. The parameters were adjusted from the left contact occupation kernels $[\bm{\mathcal{K}}^\rho_\tf{L}]_{\tf{b},\tf{b}}$ and $[\bm{\mathcal{K}}^\rho_\tf{L}]_{\tf{a},\tf{a}}$.}
\label{fig:fit}
\end{figure*}

To analyze the role of both the $\eta$ parameter and the ribbon size in the occupancy of the molecular levels, we consider fitting the occupation kernels $[\bm{\mathcal{K}}^\rho_\alpha]_{\tf{b},\tf{b}}$ and $[\bm{\mathcal{K}}^\rho_\alpha]_{\tf{a},\tf{a}}$ to a Lorentzian function. A normalized Lorentzian centered at energy $\epsilon_0$ and scaled by a factor $\kappa$ can be written as:
\begin{equation}
y(\epsilon) = \frac{\kappa}{\pi} \frac{\delta}{(\epsilon - \epsilon_0)^2 + \delta^2},
\end{equation}
where $\delta$ represents the broadening (half-width at half-maximum), and $\kappa$ sets the total area under the curve, which integrated in energy gives the maximum occupancy supported by the kernel $\mathcal{K}$, i.e.,
\begin{equation}
\kappa = \int_\Delta \tf{d}\epsilon \, \mathcal{K}(\epsilon),
\end{equation}
where $\Delta$ corresponds to an energy range containing the Kernel peak. The maximum of this peak occurs at $\epsilon = \epsilon_0$, yielding:
\begin{equation}
\max[\mathcal{K}]=\frac{\kappa}{\pi\delta} \rightarrow \delta = \frac{\int_\Delta \tf{d}\epsilon \, \mathcal{K}(\epsilon)}{\pi \, \max[\mathcal{K}]}.
\end{equation}
These relations allow the extraction of the Lorentzian parameters using the maximum value of the Kernel and its numerical integral, and these are shown in Fig~\ref{fig:fit} as functions of the $\eta$ parameter and the size of the ribbon. In panels (a) and (b) we show, in log-log scale, the peak width $\delta$ as a function of $\eta$ for graphene and Kane-Mele substrates, respectively. This is done for several ribbons widths $N = 3\ell +2$, with $\ell$ an odd number ranging from 15 to 35. For small values of $\eta$, the peak width closely approximates the physical broadening, i.e., $\delta = \Gamma_\alpha + \eta \to \Gamma_\alpha$, which reflects the actual coupling to the $\alpha$-lead. In particular, we can observe the strong dependence of $\Gamma_\alpha$ with the size of the ribbon in graphene, which is suppressed in the Kane-Mele model. As $\eta$ increases, however, the peak broadens due to the increased relaxation rate of the system. In this situation, the occupation of the molecular levels is suppressed by the interaction with the relaxation bath, i.e., $\delta \to \eta$, such that no information on the $\Gamma_\alpha$ rate is obtained. 

In Figs.~\ref{fig:fit}(c) and (d) we show the corresponding occupancies, given by the $\kappa$ prefactor. According to Section~\ref{app:wb}, this term is related with the contact rate $\Gamma_\alpha$ and $\eta$ through $\kappa = \Gamma_\alpha / (\Gamma_\alpha + \eta)$, such that in the limit $\eta \to 0$ one recovers the expected ideal occupancy.\footnote{For the shown examples, we only consider the contribution from the left contact, such that the occupation probabilities integrate to the ideal value of 1/2.} In this limit, the only relevant timescale is set by $\Gamma_\alpha$, allowing the molecular levels to reach their maximum occupancy, as there is no temporal constraint. In consequence, the occupancies become independent of the ribbon size and molecular level even in graphene substrates. 
As expected from the $\kappa$ term, for intermediate $\eta$ values the $\Gamma_\alpha$ rate plays a significant role in the occupancies, which tend to diminish when increasing the size of the ribbon in graphene. Additionally, we observe larger occupancies related with the antibonding level, which suggests an easier destabilization of the molecule under non-equilibrium conditions. Interestingly, to obtain these differences between bonding and antibonding occupancies, a reference timescale---in our case given by $\eta$---is essential. Finally, for large $\eta$ values the electrons coming from the contacts relax before reaching the molecule, so the molecular occupation is strongly reduced.

It is possible to estimate a realistic value of $\eta$ based on known decay times. For instance, electronic lifetimes due to electron–phonon interactions typically range from ${\sim}10$ ps at room temperature to ${\sim}200$ ps below 50 K \cite{strait2011}. By taking $\tau = 100$ ps as a representative value for intermediate temperatures, the corresponding broadening is $\eta = \hbar/\tau \approx 1.5 \times 10^{-5} \, \gamma_0$. Therefore, in realistic simulations, values of $\eta$ on the order of $10^{-5} \, \gamma_0$ or larger can be justified as representing physical relaxation mechanisms due to electron-phonon interactions.

\bibliographystyle{apsrev4-1_title}
\bibliography{cite}

\end{document}